\documentstyle[12pt]{article}

\newcommand{\be}{\begin{equation}}
\newcommand{\ee}{\end{equation}}
\newcommand{\bea}{\begin{eqnarray}}
\newcommand{\eea}{\end{eqnarray}}

\begin{document}
\newcommand{\nd}[1]{/\hspace{-0.5em} #1}
\begin{titlepage}
\begin{flushright}
{\bf June 2004} \\ 
SWAT-399  \\ 
hep-th/0406104 \\
\end{flushright}
\begin{centering}
\vspace{.2in}
 {\large {\bf A New Deconstruction of Little String Theory}}\\
\vspace{.4in}
Nick Dorey \\
\vspace{.4in}
Department of Physics, University of Wales Swansea \\
Singleton Park, Swansea, SA2 8PP, UK\\
\vspace{.2in}
%
%
\vspace{.4in}
{\bf Abstract} \\

\end{centering}
We present evidence for a new deconstruction 
of Little String Theory (LST). The starting point is a four-dimensional 
conformal field theory on its Higgs branch which provides a lattice 
regularization of six-dimensional gauge theory. We argue that 
the corresponding continuum limit is a 't Hooft 
large-$N$ limit of the same four-dimensional theory on an 
S-dual confining branch. The AdS/CFT correspondence is then used to study 
this limit in a controlled way. We find that the limit yields 
LST compactified to four dimensions on a torus of fixed size. The 
limiting theory also contains other 
massive and massless states which are completely 
decoupled. The proposal can be 
adapted to deconstruct Double-Scaled Little String Theory and 
provides the first example of a large-$N$ confining gauge theory 
in four dimensions with a fully tractable string theory dual. 

\end{titlepage}
\section{Introduction and Overview}
\paragraph{}
One of the more surprising outcomes of recent developments in 
string theory is the discovery of Lorentz invariant 
interacting quantum theories without gravity in spacetimes of dimension 
greater than four. In this paper we will focus on a six-dimensional 
theory known as Little String Theory (LST) which 
arises on the world volume of coincident IIB NS5 branes in a 
certain decoupling limit \cite{LST} 
(for a review see \cite{Rev1,Rev2}). The theory is non-local but reduces to a 
conventional six-dimensional non-abelian gauge theory at low energies. 
The LST corresponding to $m$ NS5 branes has low energy gauge 
group $SU(m)$. LST is interesting for a number of reasons including 
its relation to string theory on singular spacetimes and possible 
phenomenological applications. After compactification, 
LST also has an interesting relationship to 
confining gauge theories in four-dimensions 
\cite{Witten1}. In this paper, we will find a 
new and precise form of this relationship which implies that LST is fully 
equivalent to a particular large-$N$ confining gauge theory.    
\paragraph{}
Our approach to understanding LST will be based on the idea 
of {\em deconstruction} \cite{AHCG} (For related work see 
\cite{Hal}). Deconstruction provides an 
attractive way of obtaining higher dimensional theories as special limits  
of more familiar four-dimensional gauge theories. A 
deconstruction of LST using the large-$n$ limit of a four-dimensional 
quiver model with gauge group $SU(m)^{n^{2}}$ was suggested in 
\cite{AHCK}. In this paper we will discuss a related proposal \cite{AF1,D1} 
based on a different four-dimensional theory with gauge group $U(mn)$.  
In either case, the four-dimensional theory has a Higgs branch 
where the large-$n$ classical spectrum of 
massive W-bosons coincides with the Kaluza-Klein spectrum of a 
six-dimensional theory compactified on a torus. 
In fact the classical theory is equivalent to a lattice regularisation of the 
six-dimensional $SU(m)$ gauge theory which arises as the IR limit of LST. 
In both cases the proposal is that LST itself can be obtained 
as a continuum limit of this lattice theory. 
The $U(mn)$ construction of \cite{AF1,D1} actually yields a 
non-commutative generalization of lattice gauge theory. 
However, we will see that the theory becomes commutative 
in the continuum limit. 
\paragraph{}
The emergence of a six dimensional lattice theory observed in 
\cite{AHCK,AF1,D1} is based on classical arguments which are only valid at 
weak coupling. 
However, as we review below, obtaining a continuum theory necessarily 
involves taking a large-$n$, strong-coupling limit. 
The plausibility of deconstruction 
(and its usefulness) are dependent on understanding this limit.  
In this paper we will use field theory and string theory methods 
to study the continuum limit of \cite{D1} in a controlled way. 
As suggested in \cite{D1}, the first step is to use S-duality to 
reinterpret the continuum limit as a more or less conventional 
't Hooft limit of a confining gauge theory. We will then use 
the AdS/CFT correspondence to construct a holographic dual 
of the strongly coupled confining theory. The RG flow from four dimensional 
behaviour in the UV to a six-dimensional theory in the IR can then be 
exhibited directly. We are able to show 
that the proposed continuum limit does indeed yield LST, 
although the details are quite different from 
weak-coupling expectations. As an application of these ideas, we adapt our 
proposal to deconstruct double-scaled Little String Theory (DSLST). 
Weak coupling calculations in DSLST provide exact results for the large-$N$ 
glueball spectrum of the dual gauge theory. 
In the remainder of this introductory 
section we will give an overview of the main results. The details are fleshed 
out in the remaining sections.   
\paragraph{}
In the version of deconstruction suggested in \cite{AF1,D1}, 
the appearance of additional spacetime dimensions follows from a 
phenomenon which is very familiar in the context of M(atrix) theory 
\cite{BFSS}. 
We will start from ${\cal N}=4$ SUSY Yang-Mills in four-dimensions with 
gauge group $U(N)$ (for $N=mn$) 
as realized on the world volume of $N$ D3 branes in 
Type IIB string theory. 
The ${\cal N}=4$ theory contains three complex adjoint scalar fields 
denoted $\Phi_{1}$, $\Phi_{2}$ and $\Phi_{3}$. We will choose a non-zero 
background for these fields obeying, 
\begin{equation}
\Phi_{1}\Phi_{2}=\exp(-2\pi i/n)\Phi_{2}\Phi_{1}, 
\label{fterm}
\end{equation} 
and then expand the fields around this 
background. In a particular $n\rightarrow \infty$ limit, the resulting theory 
is classically equivalent to a six-dimensional $U(m)$ gauge theory 
compactified on a torus \cite{Mat,Seib}. 
In string theory, this is interpreted as the 
polarization of $N$ D3 branes into $m$ D5 branes wrapped on 
a two-dimensional torus in the transverse dimensions. 
At the classical level, 
similar considerations apply to other theories with 
sixteen supercharges including the matrix quantum mechanics 
obtained by reduction of the ${\cal N}=4$ theory to $0+1$ dimensions. 
The latter case leads to the construction of 
toroidally wrapped memebranes from D0 branes in M(atrix) theory 
\cite{WATI,sr}.           
\paragraph{}
One potential problem with this 
proceedure is that the solutions of (\ref{fterm}) are 
not vacuum states of the ${\cal N}=4$ theory and so the corresponding 
D3/D5 configuration is unstable. However, this is easily remedied by 
deforming the ${\cal N}=4$ theory. In particular, we can consider 
instead the theory with superpotential, 
\begin{equation}
{\cal W}= {\rm Tr}_{N}\left[
e^{i\frac{\beta}{2}}\Phi_{1}\Phi_{2}\Phi_{3}-e^{-i\frac{\beta}{2}}\Phi_{1}
\Phi_{3}\Phi_{2}\right]    
\label{LSsupx}
\end{equation}
Here the three complex scalars of the ${\cal N}=4$ theory have been promoted 
to ${\cal N}=1$ chiral superfields. In the ${\cal N}=1$ language, the 
theory also contains a $U(N)$ vector multiplet. As in \cite{D1}, 
we will refer to this model as the $\beta$-deformed theory. The undeformed 
${\cal N}=4$ theory corresponds to $\beta=0$. 
For $\beta=2\pi/n$, one of the resulting F-term equations coincides with 
our background condition (\ref{fterm}). The solutions of (\ref{fterm}) 
therefore yield stable vacua which preserve the ${\cal N}=1$ 
supersymmetry of the deformed theory. The non-trivial solutions 
of (\ref{fterm}) correspond to a Higgs branch of the theory where 
$U(N)$ is broken down to a $U(m)$ subgroup. 
\paragraph{}
In the string theory set-up, the deformation (\ref{LSsupx}) 
of the ${\cal N}=4$ superpotential 
corresponds to a particular background value for the Ramond-Ramond 
three-form field strength. The effect of the background field is 
to stablize the corresponding configuration of 
toroidally-wrapped D5 branes. This is entirely analogous to the 
Myers effect \cite{Myers} which causes the polarization of 
D3 branes into spherically-wrapped D5 branes in the string theory dual of 
the ${\cal N}=1^{*}$ SUSY Yang-Mills studied by Polchinski and 
Strassler \cite{PS}. 
Many features of the analysis given in this paper run 
parallel to the ${\cal N}=1^{*}$ case of \cite{PS} 
although there are also some important differences.  
\paragraph{}  
The appearance of extra dimensions in the $U(mn)$ 
$\beta$-deformed theory is also closely related to the 
conventional set-up for deconstruction based on a quiver theory with 
gauge group $SU(m)^{n^{2}}$. 
In both cases the theory has a Higgs branch with unbroken gauge group 
$U(m)$ (or $SU(m)$) and the large-$n$ spectrum of 
massive W-bosons on the Higgs branch provides two towers of Kaluza-Klein (KK) 
states. At finite-$n$, each KK tower is truncated 
in a way which corresponds to a discretization of the 
additional dimensions. In both cases the full classical action 
can be interpreted as a discretised version of 
six-dimensional gauge theory, defined 
on $R^{3,1}\times {\cal L}$ where ${\cal L}$ is an 
$n\times n$ lattice with periodic boundary conditions.  
A key difference is that expanding the $U(mn)$ theory around 
the background (\ref{fterm}) yields a {\em non-commutative} 
lattice gauge theory \cite{AF1,Nish}. 
In contrast the conventional approach to 
deconstruction based on gauge group $SU(m)^{n^{2}}$ yields an 
ordinary commutative lattice theory.    
\paragraph{}
Although the two approaches to deconstruction 
start from theories with very different gauge group and 
matter content, string theory provides an easy way to understand the 
relation between them \cite{AF1}. The quiver theory of \cite{AHCK} is realized 
in IIB string theory as the worldvolume theory of $m$ D3 branes placed 
at a ${\bf C}^{3}/{\bf Z}_{n}\times {\bf Z}_{n}$ orbifold singularity. 
On the other hand, the $\beta$-deformed theory with gauge group 
$U(mn)$ and $\beta=2\pi/n$ can be realised in string theory in (at least) 
two ways. As above we can consider $mn$ D3 branes with non-zero RR 
three-form background. An alternative construction of the same 
theory is to place $m$ D3 branes 
at a ${\bf C}^{3}/{\bf Z}_{n}\times {\bf Z}_{n}$ orbifold 
singularity with a single unit of {\em discrete torsion} \cite{MD1}. 
As in the quiver construction of \cite{AHCK}, 
the truncated towers of KK states correspond to 
fundamental strings stretched between D3 branes and their 
image points under the orbifold group. At large $n$, the orbifold 
becomes a sharp cone over an $S^{5}/{\bf Z}_{n}\times {\bf Z}_{n}$ 
base. In an appropriate scaling limit neighborhood of a point on the cone 
becomes $R^{4}\times T^{2}$ and, after T-dualizing both compact directions, 
the $m$ D3 branes become $m$ toroidally wrapped D5 branes \cite{AHCK}. 
In the $U(mn)$ construction, the limit also converts the discrete 
torsion into a background value for $B_{\rm NS}$ on $T^{2}$ which induces 
world-volume non-commutativity on the wrapped D5 branes \cite{AF1}.      
\paragraph{}
In this paper we will focus on the deconstruction of six-dimensional 
gauge theory provided by the $\beta$-deformed $U(mn)$ theory described above. 
In the following $g^{2}$ denotes the four-dimensional $U(mn)$ gauge coupling. 
It is instructive to relate the parameters of the resulting 
lattice theory to those of 
the underlying four-dimensional gauge theory. On the Higgs branch the gauge 
symmetry is broken from $U(mn)$ down to $U(m)$ at a scale
\footnote{In fact there can be several independent scales set by the 
VEVs of the different scalar fields but for simplicity we will 
supress this. For full details see Section 3 below.} 
${\rm v}$ set by the VEVs. 
The six-dimensional theory is then characterised 
by the following length scales,  
\begin{equation}
\begin{array}{cccc}
G_{6} \sim \sqrt{g^{2}n}\,{\rm v}^{-1}  & \qquad{} \,\,\,
\varepsilon \sim {\rm v}^{-1} & \qquad{}\,\, \,
\sqrt{\theta} \sim \sqrt{n}\,{\rm v}^{-1} 
& \qquad{}\,\,\,  R \sim n\,{\rm v}^{-1}
\end{array}    
\label{scalings}
\end{equation}
Here $G_{6}$ is the six-dimensional gauge coupling, 
$\varepsilon$ is the lattice spacing, $\sqrt{\theta}$ is the 
length-scale of non-commutativity and $R$ is the radius of the 
compact dimensions. For weak coupling and large-$n$, the  
four length-scales given in (\ref{scalings}) appear in 
ascending order and are well seperated. As we go from small length scales 
to larger ones, the classical theory undergoes RG flow from 
a four-dimensional conformal field theory in the UV ($l<<\varepsilon$), 
to a six dimensional non-commutative gauge theory ($\varepsilon<<l<<R$) 
and finally to an ordinary four-dimensional gauge theory in the far 
IR ($R<<l$).
\paragraph{}
The fact that the lattice spacing is much larger than the 
six-dimensional gauge coupling 
is a characteristic feature of weak-coupling deconstruction and indicates 
that the lattice theory is far from the continuum. 
Ideally we would like to find a continuum limit in 
which the non-commutivity scale $\sqrt{\theta}$ and lattice spacing 
$\varepsilon$ go to zero with 
$G_{6}$ and $R$ held fixed. Extrapolating the weak coupling 
formulae in (\ref{scalings}) indicates that this can achieved 
taking $n\rightarrow\infty$ with $g^{2}\sim n$ and ${\rm v}\sim n$. 
Naively this should yield a commutative theory with 
${\cal N}=(1,1)$ super-Poincare invariance in six dimensions 
compactified to four dimensions on a torus of fixed size. 
This continuum theory should reduce to a 
$U(m)$ gauge theory at low energies. 
As LST is the only known theory with these properties it is the natural 
candidate for the theory which arises in our proposed continuum limit.  
However, to understand this limit properly we certainly need to 
understand the quantum corrections to the classical picture 
of deconstruction described above. For example, one could easily 
imagine that the classical formula for $\varepsilon$ is corrected 
in such a way that the lattice spacing never vanishes. 
\paragraph{}
The preceeding discussion indicates that 
the interesting questions about the existence and nature of 
a continuum limit are hidden in the strongly-coupled dynamics 
of the four-dimensional gauge theory. 
Several remarkable properties of the quantum theory with superpotential 
(\ref{LSsupx}) will allow us to make progress in answering these questions. 
For other relevant 
work on this model see \cite{BL1}.  
The first point is that 
the parameter $\beta$ corresponds to an {\em exactly marginal} deformation 
of the ${\cal N}=4$ theory \cite{LS1}. 
The deformation parameter 
$\beta$ does not run and therefore parametrizes a  
family of ${\cal N}=1$ superconformal field theories. 
The Higgs branch which appears 
classically for $\beta=2\pi/n$ persists in the quantum theory for all 
values of the gauge coupling \cite{D1}. Another remarkable property, 
uncovered in 
\cite{D1,DHK}, is that the the theory has an exact electric-magnetic duality 
extending that of the ${\cal N}=4$ theory. As in the ${\cal N}=4$ theory 
the duality inverts the coupling constant: $g^{2}\rightarrow 
\tilde{g}^{2}=16 \pi^{2}/g^{2}$. The transformation also acts non-trivially 
on the deformation parameter taking the theory with $\beta=2\pi/n$ 
described above to a dual theory with 
$\tilde{\beta}=8\pi i/\tilde{g}^{2}n$. The theory with this imaginary 
value of the deformation parameter has no Higgs branch but instead 
has a new branch, invisible classically, 
where the $U(N)$ gauge symmetry is confined down to $U(m)$ at a scale
\footnote{As explained in \cite{D1}, 
the factor of $\tilde{g}^{2}/4\pi$ in the relation between the 
scales ${\rm v}$ and $\tilde{\rm v}$ comes from the transformation 
properties of chiral operators under S-duality.} 
$\tilde{\rm v}=\tilde{g}^{2}v/4\pi$. The physics of this phase is 
discussed in \cite{D1}. The candidate 
continuum limit of the Higgs branch theory can now be reinterpreted 
as a limit of the theory on this new quantum confining branch. 
Specifically we must take now take the limit $n\rightarrow \infty$, 
$\tilde{g}^{2}\rightarrow 0$ with $\tilde{g}^{2}n$ and 
$\tilde{\rm v}$ held fixed. 
\paragraph{}
Interestingly the S-dual continuum limit is something 
quite familiar: a 't Hooft large-$N$ limit\footnote{Throughout this 
paper we have $N=mn$ and the large-$N$ limit we consider 
corresponds to taking $n\rightarrow\infty$ with $m$ fixed.} 
of gauge theory in a 
confining phase, although the confinement is only partial. This is 
a limit in which we expect Yang-Mills theory to exhibit string-like 
behaviour. On the other hand LST, which we will claim arises 
in this limit, is a non-critical theory of closed strings. 
It is natural to suspect that the Little String 
is one and the same as the confining string in large-$n$ gauge theory.      
In the final Section of the paper, we will find a region of parameter space 
where this correspondence can be made quite precise.  
\paragraph{}
As promised above our main goal is to find a weakly-coupled dual 
description in which we can study the proposed continuum limit.    
We will accomplish this by applying yet another 
duality to the confining phase theory. In particular, 
when $\tilde{g}^{2}n>>1$, we have $|\tilde{\beta}|=8\pi/\tilde{g}^{2}n<<1$. 
The resulting theory is then a small deformation of 
${\cal N}=4$ SUSY Yang-Mills at large-$N$, with large 't Hooft coupling 
$\tilde{g}^{2}N>>1$. The conformally invariant vacuum of 
the theory therefore has a reliable 
description in IIB supergravity as a small deformation of 
$AdS_{5}\times S^{5}$ \cite{Mal}. 
The deformation in question involves the introduction 
of non-zero NS three-form flux on the boundary of $AdS_{5}$ and has been 
worked out explicitly in \cite{GP,AKY}. 
\paragraph{}
Following the ideas of 
Polchinski and Strassler \cite{PS}, we can also find AdS duals for the 
various Higgs and confining phase ground states of this theory by introducing 
wrapped five-branes embedded in this geometry. 
In particular, the string dual of the 
confining vacuum involves $m$ NS fivebranes wrapped on a 
two-dimensional torus $T^{2}\subset S^{5}$. The fivebranes are located at 
fixed radial distance. The $N$ D3 branes also expand to lie on the same 
toroidal shell and the resulting geometry is warped accordingly 
As in the ${\cal N}=1^{*}$ case, we will find an approximate 
supergravity solution (valid for $\tilde{g}^{2}n>>m$) 
corresponding to this brane configuration. 
\paragraph{}
Far from the branes, the dual geometry of the confining phase 
asymptotes to $AdS_{5}\times S^{5}$ deformed by a background NS 
threeform flux. This corresponds to the strongly-coupled four-dimensional 
superconformal fixed point which controls the UV behaviour 
of the dual field theory. As we approach the branes, 
the solution makes a smooth transition to the near-horizon geometry 
$m$ NS5 branes. The spectrum of theory includes the $U(m)$  
gauge fields living on the six-dimensional world volume of the 
NS5 branes. The SUGRA solution fixes the six-dimensional 
parameters in terms of the four-dimensional ones. We can then compare 
these strong coupling results from a naive extrapolation of the classical 
formulae (\ref{scalings}). We find that the 
strong and weak coupling results for the six-dimensional gauge coupling 
$G_{6}$ and the compactification radius $R$ agree. In addition the  
low energy six-dimensional gauge theory in this regime is commutative 
as expected.    
\paragraph{}
The dual geometry described above encodes the RG flow from 
a four dimensional CFT in the UV to a six-dimensional gauge theory in the IR. In particular it determines the mass scale at which this 
transition takes place. We find that 
the strongly-coupled theory behaves like a four-dimensional CFT 
above the scale $\Lambda \sim\tilde{\rm v}/\sqrt{\tilde{g}^{2}mn}$. 
At weak coupling the corresponding scale is set by the inverse 
lattice spacing $\varepsilon^{-1}$. 
Here we find a significant discrepancy between the strong and weak coupling 
results. In particular, the strong-coupling scale $\Lambda$ remains 
fixed in our proposed continuum limit. We conclude that the theory does not 
recover six-dimensional Lorentz invariance in this limit.  
\paragraph{}
Despite this negative result, the dynamics of the 
strongly-coupled theory simplifies in an interesting way in the large-$n$ 
't Hooft limit discussed above. The supergravity dual of the confining phase 
involves $m$ NS fivebranes embedded in a geometry which is asymptotically 
AdS. The 't Hooft limit involves taking the asymptotic string coupling 
$\tilde{g}_{s}=\tilde{g}^{2}/4\pi$ to zero. For NS5 branes embedded in 
asymptotically flat space a similar limit decouples the degrees of freedom 
on the fivebranes. The resulting theory is precisely Little 
String Theory. We will show that the 't Hooft limit has  
the same effect in the present case. The effective 
string coupling goes to zero everywhere except in a region very close 
to the fivebranes. In this region the solution 
coincides with the near horizon geometry of $m$ toroidally 
wrapped NS fivebranes with no other SUGRA fields turned on. This geometry 
is holographically dual to LST \cite{holog}. 
\paragraph{}
The decoupling described above has a simple interpretation in the dual 
confining gauge theory. 
The Hilbert space of the large-$N$ theory contains a sector where states form 
towers of Kaluza-Klein modes. These states and their interactions respect 
six-dimensional Lorentz invariance.   
The theory also contain another 
sector of states which badly violate the six-dimensional Lorentz invariance. 
At large but finite $n$ the two sectors are weakly-coupled to each other.  
In the 't Hooft large-$N$ limit, the states in the four-dimensional sector 
retain finite masses but decouple completely both from each other and from 
the six-dimensional sector. The six-dimensional sector remains 
interacting and is exactly Little String Theory. Our main result can 
therefore be summarised as follows:  
\paragraph{}
{\bf Result} We consider the $\beta$-deformation of ${\cal N}=4$ SUSY 
Yang-Mills with gauge group $U(mn)$, gauge coupling $\tilde{g}^{2}$ and 
deformation parameter $\tilde{\beta}=8\pi^{2}i/\tilde{g}^{2}n$. The theory 
has a vacuum where the $U(mn)$ gauge group is confined down to a $U(m)$ 
subgroup at scale $\tilde{\rm v}$. In the limit $n\rightarrow \infty$, 
$\tilde{g}^{2}\rightarrow 0$ with $\tilde{g}^{2}n$, $m$ and 
$\tilde{\rm v}$ fixed, the interacting sector of the theory is equivalent 
to Little String Theory with low energy gauge coupling 
$G_{6}\sim \sqrt{\tilde{g}^{2}n}\tilde{\rm v}^{-1}$ 
compactified on a torus of radii $R\sim
\tilde{g}^{2}n\tilde{\rm v}^{-1}$. 
\paragraph{}
The precise statement of this result is given in Section 8 below.  
The supergravity analysis which establishes this result is valid 
for $\tilde{g}^{2}n>>m$ 
where the compactification torus is large compared to the gauge coupling 
of LST. We will also conjecture that the result holds for more 
general values of $\tilde{g}^{2}n$ although the evidence for this is 
limited.            
\paragraph{}
The results described in this paper reveal a new aspect of the duality between 
large-$N$ confining gauge theory and string theory. As with other dualities, 
it usefulness depends on identifying regions in parameter space where 
calculations can be performed on (at least) one side of the 
correspondence. On the string theory side, the theory on $m$ NS5 branes 
has an interesting double-scaling limit \cite{GK,AGK}. The resulting so-called 
Double-Scaled Little String Theory is holographically dual to a 
weakly-coupled IIB background with an exactly solvable worldsheet  
conformal field theory. In Section 7, we adapt the results described above 
to deconstruct DSLST. The dual gauge theory is realised in a phase where 
$U(mn)$ is confined down to $U(1)^{m}$. The weak coupling regime of DSLST 
corresponds to a regime where all states charged under the low-energy gauge 
symmetry are very massive. We argue that the standard large-$N$ scaling 
arguments for confining gauge theories apply in this regime and should 
lead to an infinite tower of weakly interacting glueball states with an 
S-matrix exhibiting Regge behaviour. We find 
that DSLST provides, for the first time, an 
analytic description of this phenomenon in terms of weakly-coupled 
closed string theory. 
\paragraph{}
Another aspect of the duality described above is that it provides a 
new non-perturbative definition of LST and therefore, via holography, of 
linear dilaton backgrounds. Although we will not develop this viewpoint much 
in the present paper, we make some 
preliminary comments in the final section and 
hope to return to this topic in future work. The rest of the paper is  
organised as follows. 
In Section 2, we review some of the properties 
of LST and also introduce a more general class of fivebrane theories 
which reduce to non-commutative gauge theories at low energy. 
Section 3 reviews the $\beta$-deformation of ${\cal N}=4$ SUSY 
Yang-Mills including some of the main results of \cite{D1}. 
In Section 4 we review the basic idea of deconstruction at the classical 
level. In Section 5 we take a first look at the continuum limit. 
Section 6 is devoted to constructing a string theory dual of the 
Higgs branch vacuum and Section 7 reviews the corresponding dual for the 
confining phase. In Section 8 we discuss the continuum limit and formulate a 
precise version of the deconstruction conjecture. In Section 9 we adapt our 
results to the case of DSLST described above. 
Some calculational details from Sections 6 and 9 are relegated to 
Appendices A and B respectively.            
 
\section{LST and its Non-Commutative Cousins}
\paragraph{}
The basic definition of Little String Theory (LST) is as a decoupling 
limit of the worldvolume theory of fivebranes in Type II string theory. 
We will start from $m$ parallel D5 branes of the Type IIB theory with 
string coupling $g_{s}$ and squared string length $\alpha'$. 
At low-energy the theory on the world volume reduces to a six dimensional 
$U(m)$ gauge theory with ${\cal N}=(1,1)$ supersymmetry and 
gauge coupling $\hat{G}_{6}=\sqrt{16\pi^{3}\alpha'g_{s}}$. 
As for the lower dimensional Dirichlet branes we can try to take a limit 
which isolates the worldvolume theory. To decouple the excited modes of 
the open strings ending on the D5's we need to take the limit 
$\alpha'\rightarrow 0$. If we also try to keep the low-energy gauge coupling 
fixed we are forced to simultaneously take the limit 
$g_{s}\rightarrow \infty$. 
\paragraph{}
We can understand this limit better using the S-duality transformation of the 
IIB theory which acts on the parameters as,  
\begin{eqnarray}
g_{s}\rightarrow \tilde{g}_{s}=\frac{1}{g_{s}} \qquad{} &  \qquad{} 
& \alpha' \rightarrow \tilde{\alpha}'=g_{s}\alpha' 
\label{sdual0}
\end{eqnarray}  
This transformation maps the $m$ D5 branes we started with to a 
configuration of $m$ parallel NS5 branes. In terms of the S-dual variables, 
the low energy gauge coupling 
is $\hat{G}_{6}=\sqrt{16\pi^{3}\tilde{\alpha}'}$ and the decoupling limit 
becomes simply $\tilde{g}_{s}\rightarrow 0$ with $\tilde{\alpha}'$ 
held fixed. In this limit the ten-dimensional Planck length goes to zero 
and the theory on the branes is decoupled from gravity. We will now briefly 
review the basic properties of the resulting theory:  
\paragraph{}
{\bf 1:} The IIB LST has six-dimensional ${\cal N}=(1,1)$ 
super-Poincare invariance. The theory also has an exact $SO(4)$ R-symmetry 
corresponding to rotations of the four transverse directions of the branes. 
\paragraph{}
{\bf 2:} Apart from the integer $m$, the 
theory has a single parameter, a characteristic mass scale which 
in our conventions is 
$\hat{M}_{s}=1/\sqrt{16\pi^{3}\tilde{\alpha}'}$. 
At energies below the scale $\hat{M}_{s}/\sqrt{m}$, 
it reduces to ${\cal N}=(1,1)$ supersymmetric Yang-Mills theory 
in six dimensions with gauge coupling $\hat{G}_{6}=1/\hat{M}_{s}$. 
The resulting low-energy gauge group 
is $U(m)$. The $U(1)$ vector multiplet 
corresponding to the center of $U(m)$ is completely decoupled and the 
remaining interacting sector of the theory has low-energy 
gauge-group $SU(m)$     
\paragraph{}
{\bf 3:} The theory has a moduli space ${\rm Sym}^{m}{R}^{4}$ corresponding to 
the Coulomb branch of the low-energy gauge theory. Away from the origin 
the $U(m)$ gauge symmetry is broken to $U(1)^{m}$. 
\paragraph{}
{\bf 4:} In addition to the massless gauge multiplet, 
the spectrum of the $U(m)$ theory contains BPS saturated 
strings with tension $\hat{T}=8\pi^{2}\hat{M}_{s}^{2}$. Roughly speaking 
these strings correspond to bound states of the IIB string with 
the NS5 brane. 
\paragraph{}
{\bf 5:} When the theory is compactified on a circle the existence of string 
winding modes leads to an exact T-duality relating IIB LST and the 
corresponding LST on the IIA NS fivebrane. This property indicates that LST 
is not a local quantum field theory.      
\paragraph{} 
It will be useful to introduce a slightly more 
general class of six-dimensional theories first considered in 
\cite{HOSJ}. The theories in question reduce to {\em non-commutative} 
gauge theories with ${\cal N}=(1,1)$ SUSY in the IR. They are 
obtained by taking an appropriate decoupling 
limit on the worldvolume of Type IIB D5 branes with a non-zero 
background for $B_{\rm NS}$. 
These theories played an important role in the 
analysis of the ${\cal N}=1^{*}$ theory in \cite{PS} .  
They will also enter in our analysis of deconstruction although we 
emphasize that our final results concern conventional commutative LST.  
\paragraph{}
We start from a configuration consisting of $m$ flat D5 branes extended in the 
$0,1,2,3,4,5$ directions of $R^{9,1}$. In addition we will introduce a 
constant two-form potential in the $y_{4}$-$y_{5}$ plane, 
\begin{equation}
B_{\rm NS}=\tan\varphi \, dy_{4}\wedge dy_{5}   
\label{bns2}
\end{equation}   
The supergravity solution for this configuration was given in 
\cite{HOSJ}. The string frame metric, RR four-form, 
NS two-form, and dilaton fields read,   
\begin{eqnarray}
ds^{2} \,\,  & = & f^{-\frac{1}{2}}(u)\left[\eta_{\mu\nu}dy^{\mu}dy^{\nu} + 
h(u)\left(dy_{4}^{2}+dy_{5}^{2}\right)\right]+ 
\alpha'^{2}f^{\frac{1}{2}}(u)\left[ du^{2}+u^{2}d\Omega_{3}^{2}\right] 
\nonumber \\
B_{\rm NS} \,\, & = & \tan\varphi f^{-1}(u)h(u)\,\,  
dy_{4}\wedge dy_{5} \nonumber \\ 
\exp(2\phi) & = & g_{s}^{2} f^{-1}(u)h(u) \nonumber \\
\chi_{4} \,\, & = & \frac{1}{g_{s}} \sin\varphi f^{-1}(u)\,\,  
dy_{0}\wedge dy_{1}\wedge dy_{2}\wedge dy_{3} \nonumber \\ 
\label{d5soln}
\end{eqnarray}
where $u=s/\alpha'$ is a rescaling of the distance, $s=
\sqrt{y_{6}^{2}+\ldots+y_{9}^{2}}$, from the branes in the four transverse 
directions. 
The functions of the radial variable $u$ 
appearing in the solution are given by, 
\begin{eqnarray}   
f(u) & = &  1 + \frac{R^{2}}{\alpha'^{2}u^{2}} \nonumber \\
h^{-1}(u) & = & \sin^{2}\varphi\, f^{-1}(u) + \cos^{2}\varphi \nonumber \\ 
\label{functions}
\end{eqnarray}
with $R^{2} = g_{s}\alpha'm/\cos \varphi$. The self-duality of the 
RR five-form field strength is imposed by setting 
$F_{5}  =  d\chi_{4}+ \star d\chi_{4}$. There are also $m$ units of 
RR three-form flux $F_{(3)}$ through a three-sphere surrounding the branes 
but the explicit form of this field will not be needed in the following. 
\paragraph{}
We will now take a decoupling limit of the sort introduced by Seiberg and 
Witten \cite{SWNC}. Specifically we take the $\alpha'\rightarrow 0$ limit 
with $g_{s}$ and $b=\alpha'\tan\varphi$ held fixed. The SW limit also 
requires us to rescale the coordinates $y_{5}$ and $y_{6}$ 
in the directions of 
non-zero $B_{\rm NS}$ field. Specifically we define 
$\bar{y}_{4}=(b/\alpha')y_{4}$ and $\bar{y}_{5}=(b/\alpha')y_{5}$ and 
hold $\bar{y}_{4}$, $\bar{y}_{5}$ and the rescaled radial coordinate $u$ fixed 
as $\alpha'\rightarrow 0$. Standard arguments based on the quantisation of 
open strings ending on the $D5$ brane tell us that the low-energy theory 
on the brane is maximally supersymmetric Yang-Mills theory in $5+1$ dimensions 
with gauge coupling $G_{6}^{2}=16\pi^{3}g_{s}b$. The theory 
has non-commutativity in the $\bar{y}_{4}$-$\bar{y}_{5}$ plane:  
$[\bar{y}_{4},\bar{y}_{5}]=2\pi i\,b$. 
\paragraph{}
In the limit of interest, the excited modes of the open string decouple 
as $\alpha'$ goes to zero. 
As the ten-dimensional Planck mass 
goes to infinity in this limit we also expect the theory 
to decouple from gravity\footnote{The issue of whether 
gravity really decouples in such a limit is often subtle. 
Whether or not this is the case for the family of five-brane 
worldvolume theories considered in this section will not affect 
the main conclusions of the paper.}. Thus the final result is a 
non-gravitational theory which reduces to six-dimensional 
non-commutative $U(m)$ gauge theory at low energies. For brevity 
we will denote the resulting decoupled world-volume theory as 
${\cal T}[M_{s},g_{s}]$. This theory has a 
characteristic mass scale $M_{s}=1/\sqrt{16\pi^{3}g_{s}b}=1/G_{6}$. 
In addition to 
the elementary excitations of the gauge fields, the theory also contains 
BPS strings of tension $T=8\pi^{2}M_{s}^{2}$ 
corresponding to non-commutative Yang-Mills instantons 
embedded in six dimensions. 
In string theory language, these correspond to D-strings 
bound to the D5 branes.
\paragraph{}
Note that  
string coupling $g_{s}$ is held fixed in the decoupling limit and remains as 
an additional parameter of the theory. This is different 
from the case of D5 branes without background $B_{\rm NS}$ considered above 
where the only 
possible decoupling limit involves taking the limit $g_{s}\rightarrow \infty$ 
leading to the conventional definition of LST. 
The energy scale above which the non-commutativity in the 
$\bar{y}_{4}$-$\bar{y}_{5}$ plane becomes important is\footnote{More 
precisely this formula for $M_{\rm NC}$ applies only when 
the corresponding six-dimensional gauge theory is weakly coupled. 
This is the case provided $M_{\rm NC}<<M_{s}/\sqrt{m}$.} 
$M_{\rm NC}=1/\sqrt{2\pi b}=\sqrt{8\pi^{2}g_{s}}M_{s}$. The role of the 
extra parameter $g_{s}$ is therefore to set the ratio 
$M_{\rm NC}/M_{s}$. This suggests that non-commutativity should 
disappear in the strong coupling limit $g_{s}\rightarrow \infty$. 
We will study this limit more carefully below and see that it 
yields ordinary (ie commutative) LST.
\paragraph{}
We will now consider different regimes 
in which the decoupled fivebrane theory has a weakly coupled effective 
description.  
Like any non-abelian gauge theory in six dimensions the low-energy 
gauge theory description of ${\cal T}[M_{s},g_{s}]$   
becomes strongly coupled in the UV at scales above that set by 
the inverse 't Hooft coupling $M_{s}/\sqrt{m}$ and perturbation theory breaks 
down. To understand the behaviour of the theory in this regime we can 
consider instead the dual gravitational background.      
In the decoupling limit discussed above the D5 solution becomes,
\begin{eqnarray}
ds^{2} \,\,  & = & \alpha' \left(\frac{au}{b}\right)
\left[\eta_{\mu\nu}dy^{\mu}dy^{\nu} + 
\bar{h}(u)\left(d\bar{y}_{4}^{2}+d\bar{y}_{5}^{2}\right)\right]+ 
\alpha'\left(\frac{b}{au}\right)
\left[ du^{2}+u^{2}d\Omega_{3}^{2}\right] 
\nonumber \\
B_{\rm NS} \,\, & = & \frac{\alpha'}{b}\, a^{2}u^{2}\bar{h}(u)\,\,d\bar{y}_{4}
\wedge d\bar{y}_{5} \nonumber \\ 
\exp(2\phi) & = & g_{s}^{2}a^{2}u^{2}\bar{h}(u) 
\nonumber \\
\chi_{4} \,\, & = & \frac{1}{g_{s}} \left(\frac{\alpha'^{2}a^{2}u^{2}}
{b^{2}}\right)\,\,  
dy_{0}\wedge dy_{1}\wedge dy_{2}\wedge dy_{3} \nonumber \\
\label{d5soln2}
\end{eqnarray}     
where, 
\begin{eqnarray}
\bar{h}(u)=\frac{1}{1+a^{2}u^{2}} & \qquad{} & {\rm with}\,\,\,\, a^{2}
=\frac{b}{g_{s}m} \nonumber \\
\label{h2}
\end{eqnarray}
As before there are also $m$ units of RR threeform flux through an 
$S^{3}$ surrounding the D5 branes which we have not shown explicitly.   
\paragraph{}
The rescaled radial coordinate 
$u=s/\alpha'$ corresponds to the energy of streched fundamental strings 
ending on a probe D5 brane placed at fixed radial position $s$. 
By analogy with the UV/IR correspondence of 
more familiar conformal examples of holography, it is tempting to 
interpret dependence on the coordinate $u$ as RG flow in the 
worldvolume theory. As emphasised in \cite{peet}, there is no direct 
generalization of the UV/IR correspondence for the near horizon geometry 
fivebranes. The energy scale corresponding to a fixed value of $u$ 
depends on the process considered.            
\paragraph{}
We can now determine when the dual supergravity background 
(\ref{d5soln2}) is weakly coupled. 
The validity of tree-level string theory in the above background 
requires that the effective string coupling $e^{\phi}$ is small everywhere. 
The dilaton solution in (\ref{d5soln2}) 
shows that $e^{\phi}$ is a monotonically increasing function of the 
radial coordinate $u$ which tends to the constant value $g_{s}$. 
Hence provided $g_{s}<<1$ the dilaton is small everywhere and string loops are 
supressed. The supergravity approximation to string theory 
is valid as long as the 
curvature of the solution is small. This is the case provided that, 
\begin{equation}
u >>  \frac{1}{\sqrt{g_{s}bm}}= \frac{M_{s}}{\sqrt{m}} 
\label{valid1}
\end{equation}
For smaller values of $u$ the curvature becomes large and the 
theory is better described by weakly-coupled six-dimensional non-commutative 
Yang-Mills theory discussed above. With a naive interpretation of $u$ 
as an energy scale in the worldvolume theory, this matches the fact 
that the low-energy gauge theory becomes strongly-coupled in the UV 
at the scale $M_{s}/\sqrt{m}$ set by the inverse 't Hooft coupling. 
In other words the domains of validity of 
supergravity and Yang-Mills theory are exactly complimentary. 
\paragraph{}
As the D5 brane configuration we started from is BPS saturated  
and exists for all values of the string coupling 
we will assume that the world volume theory ${\cal T}[g_{s},M_{s}]$ also 
makes sense  for all values of $g_{s}$. Another regime where we can hope to 
study the theory succesfully is that of very large $g_{s}$ which we can 
map to weak coupling via the S-duality of the IIB theory. Specifically 
the fields and parameters transform as; 
\begin{eqnarray}
g_{s}\rightarrow \tilde{g}_{s}=\frac{1}{g_{s}} \qquad{} &  \qquad{} 
& \alpha' \rightarrow \tilde{\alpha}'=g_{s}\alpha' \nonumber \\
\exp(\phi)\rightarrow \exp(\tilde{\phi})=\exp(-\phi) & & 
ds^{2}\rightarrow d\tilde{s}^{2}=g_{s}\exp(-\phi)ds^{2} \nonumber \\
\label{2bsdualb}
\end{eqnarray}
Under these transformations the background (\ref{d5soln2}) gets mapped to, 
\begin{eqnarray}
d\tilde{s}^{2} \,\,  & = & 
\left(\frac{\tilde{\alpha}'\tilde{g}_{s}}{b}\right)\bar{h}^{-\frac{1}{2}}(u)
\left[\eta_{\mu\nu}dy^{\mu}dy^{\nu} + 
\bar{h}(u)\left(d\bar{y}_{4}^{2}+d\bar{y}_{5}^{2}\right)+ 
\left(\frac{b^{2}}{a^{2}u^{2}}\right)
\left( du^{2}+u^{2}d\Omega_{3}^{2}\right)\right] 
\nonumber \\
B_{\rm RR} \,\, & = & \tilde{g}_{s}\tilde{\alpha}'\,
\left( \frac{a^{2}u^{2}}{b}\right)\,
\bar{h}(u)\,\,\, d\bar{y}_{4}\wedge d\bar{y}_{5} \nonumber \\ 
\exp(2\tilde{\phi}) & = & \left(\frac{\tilde{g}^{2}_{s}}{a^{2}u^{2}}\right)
\bar{h}^{-1}(u) 
\nonumber \\ 
\chi_{4} \,\, & = & \frac{1}{\tilde{g}_{s}} \left(\frac{
\tilde{\alpha'}^{2}\tilde{g}^{2}a^{2}u^{2}}
{b^{2}}\right)\,\,  
dy_{0}\wedge dy_{1}\wedge dy_{2}\wedge dy_{3} \nonumber \\
\label{ns5soln2}
\end{eqnarray}     
The solution also has $m$ units of flux for the NS 
three-form field stength through an $S^{3}$ surrounding the branes.  
In terms of the new variables, the radial coordinate is 
$u=r/\tilde{g}_{s}\tilde{\alpha}'$ which corresponds to the energy of 
stretched D-strings ending on a probe NS5 brane placed a distance $r$ 
from the other branes. The functions of $u$ appearing in (\ref{ns5soln2}) 
are,   
\begin{eqnarray}
\bar{h}(u)=\frac{1}{1+a^{2}u^{2}} \qquad{} & \qquad{} & {\rm with}\,\,\, a^{2}
=\frac{b\tilde{g}_{s}}{m} \nonumber \\
\label{h5}
\end{eqnarray}
\paragraph{}
As before we can determine the regime where the dual supergravity background 
is weakly coupled. 
The validity of tree-level string theory requires that the effective 
string coupling $e^{\tilde{\phi}}$ is small. 
In the background (\ref{ns5soln2}) the dilaton is a monotonically decreasing 
function of the radial coordinate 
$u$ which tends to a constant value 
$e^{\tilde{\phi}_{\infty}}=\tilde{g}_{s}$ at $u=\infty$. 
Hence for $\tilde{g}_{s}<<1$ tree-level string theory is valid for 
large-$u$ which 
(roughly) corresponds to the UV region of the the theory on the brane. 
The dilaton becomes order one at a scale,             
\begin{equation}
u\sim \sqrt{\frac{m\tilde{g}_{s}}{b}}=\sqrt{m}M_{s}
\label{dilone}
\end{equation}
For values of $u$ below this scale we should undo the S-duality 
transformation and return to the D5-brane background. 
The supergravity approximation is valid as long as the curvature of the 
solution is small. This is the case provided that,
\begin{eqnarray}
u & >> & \sqrt{\frac{\tilde{g}_{s}}{bm}}= \frac{M_{s}}{\sqrt{m}} \nonumber \\
\label{validb1}
\end{eqnarray}  
\paragraph{}
We find that the SUGRA solution undergoes an 
important transition at the scale, 
\begin{eqnarray}
u & \sim & a^{-1}=\frac{\sqrt{m}}{\tilde{g}_{s}}M_{s} \nonumber \\
\label{valid2b}
\end{eqnarray} 
Notice that for 
$\tilde{g}_{s}<<1$ this scale is higher than the scales 
(\ref{dilone},\ref{validb1}) 
indicating that the crossover scale lies inside the regime of validity of 
classical supergravity. When $u<<a^{-1}$ we have $h(u)\simeq 1$ and the 
familiar throat region of the NS5 brane solution appears. In this region, 
the solution can be written as,
\begin{eqnarray}
d\tilde{s}^{2} \,\,  & = & \eta_{AB}dY^{A}dY^{B} + 
\tilde{\alpha}'m \frac{d\hat{u}^{2}}{\hat{u}^{2}}+ 
\tilde{\alpha}'m d\Omega_{3}^{2} 
\nonumber \\
\exp(2\tilde{\phi}) & = & \frac{m}{\tilde{\alpha}'}\, \frac{1}{\hat{u}^{2}} 
\nonumber \\
\label{ns5soln3}
\end{eqnarray}
where $\eta_{AB}$ with $A,B=0,1,\ldots,5$ is the standard flat metric 
on six-dimensional Minkowski space. 
Note that we have defined rescaled coordinates along the brane as 
$Y_{A}=(M_{s}/\hat{M}_{s})y_{\mu}$ for $A=\mu=0,1,2,3$, 
and  $Y_{A}=(M_{s}/\hat{M}_{s}) 
\bar{y}_{4,5}$ for $A=4,5$. 
We have also rescaled 
the radial coordinate as $\hat{u}=
(\hat{M}_{s}/M_{s})u$ where $\hat{M}_{s}=1/\sqrt{16\pi^{3}\tilde{\alpha}'}$. 
\paragraph{}
As before we have $m$ units of NS 3-form flux through an 
$S^{3}$ surrounding the branes. However, all 
the other fields of IIB supergravity 
go to zero rapidly for $u<<a^{-1}$. Importantly, the 
solution in this region therefore has no non-zero RR field strengths. 
In fact the solution (\ref{ns5soln3}), is simply the near 
horizon geometry of $m$ NS5 branes with no additional fields turned on. 
The geometry in 
question is known to be holographically dual to ordinary 
IIB  Little String Theory with low-energy gauge group $SU(m)$ \cite{holog}. 
As above, the dynamics of the latter theory is characterised by the mass scale 
$\hat{M}_{s}=1/\sqrt{16\pi^{3}\tilde{\alpha}'}$. 
The rescaling of the coordinates described mean that 
lengths and energies measured in LST are not the same as those of 
the theory ${\cal T}[g_{s},M_{s}]$ but differ by appropriate powers of 
$M_{s}/\hat{M}_{s}$. The net effect of this rescaling is to 
replace the mass parameter $\hat{M}_{s}$ of LST with the mass parameter 
$M_{s}$ of the theory worldvolume theory ${\cal T}[g_{s},M_{s}]$.         
\paragraph{}
Thus we find that, when $g_{s}>>1$, the holographic description of 
our non-commutative world-volume theory ${\cal T}[M_{s},g_{s}]$ 
simplifies for small values of the radial coordinate $u$. 
States localised far down the throat at $u<<a^{-1}$ are effectively 
describe by LST with low energy gauge group $SU(m)$ and mass 
parameter $M_{s}$. Roughly speaking this is a 
strong coupling analog of the fact that non-commutative gauge theory 
reduces to its commutative counter part at low energies. However  
the analogy is imprecise because, as mentioned above, 
the relation between the radial coordinate and 
energy in the boundary theory is ambiguous. 
\paragraph{}       
It will be useful to understand what 
happens if we now take the limit $g_{s}\rightarrow \infty$ with the scale 
$M_{s}$ fixed. As the S-dual coupling $\tilde{g}_{s}$ goes to zero the 
string coupling vanishes every where except for values of the radial 
coordinate $u$ satisfying $u<<a^{-1}$ for which the geometry is given by 
the NS5 brane throat solution (\ref{ns5soln3}). Thus the states located 
in the throat region retain finite interactions while all other states 
decouple. The upshot is that the resulting theory 
${\cal T}[M_{s},g_{s}\rightarrow \infty]$ includes the 
IIB LST with low-energy gauge group $SU(m)$ along with other states which 
are completely decoupled. Among the decoupled states are the $U(1)$ 
gauge boson corresponding to the center of the original gauge group 
$U(m)$ and its ${\cal N}=(1,1)$ superpartners. 

\section{Review of the $\beta$-deformation}

\paragraph{}
In this Section we review the key features of 
the $\beta$-deformation of ${\cal N}=4$ SUSY Yang-Mills 
theory studied in \cite{D1}. 
In terms of ${\cal N}=1$ superfields, this 
theory contains a $U(N)$ vector multiplet $V$ and three chiral multiplets 
$\Phi_{i}$, with $i=1,2,3$, in the adjoint representation of the gauge group. 
The classical superpotential is given as,     
\begin{equation}
{\cal W}= i\kappa {\rm Tr}_{N}\left(\Phi_{1}[\Phi_{2},\Phi_{3}]_{\beta}\right)
\label{LSsup}
\end{equation}
where, 
\begin{equation}
[\Phi_{i},\Phi_{j}]_{\beta}=\exp\left(i\frac{\beta}{2}\right)
\Phi_{i}\Phi_{j}- \exp\left(-i\frac{\beta}{2}\right)
\Phi_{j}\Phi_{i}
\label{deformed}
\end{equation}
The ${\cal N}=4$ theory is recovered for $\beta=0$ and $\kappa=1$. 
Apart from the complex parameters $\beta$ and $\kappa$ appearing in the 
superpotential, the theory also depends on the complexified gauge coupling 
$\tau=4\pi i/g^{2}+\vartheta/2\pi$. We will now review the classical and 
quantum properties of this theory in turn.    
\subsection{The classical theory}
\paragraph{}
In the classical theory the complex parameter $\kappa$ has no effect 
and it can be set to one. In contrast the classical vacuum structure 
of the theory depends strongly on the deformation parameter $\beta$. 
In particular, new Higgs branches appear at special values of $\beta$.  
We will focus on one of these branches which occurs when $\beta=2\pi/n$ where 
$n$ is a divisor of $N=mn$. In this case the theory 
has a Higgs branch (denoted ${\cal H}_{m}$ in \cite{D1}) 
where the gauge symmetry is broken down to a 
$U(m)$ subgroup. The scalar expectation values on this branch are given as, 
\begin{equation}
\begin{array}{ccc} \langle \Phi_{1}\rangle = \alpha_{1}
I_{(m)}\otimes U_{(n)} &
\qquad{}  
\langle  \Phi_{2}\rangle = \alpha_{2}I_{(m)}\otimes V_{(n)} & \qquad{}   
\langle \Phi_{3}\rangle = \alpha_{3}I_{(m)}\otimes V^{\dagger}_{(n)} 
U^{\dagger}_{(n)} \end{array}
\label{vacm}
\end{equation}
where $\alpha_{1}$, $\alpha_{2}$ and $\alpha_{3}$ are complex numbers. Here 
$I_{(m)}$ denotes the $m\times m$ unit matrix, and $U_{(n)}$ and $V_{(n)}$ 
are the $n\times n$ clock and shift matrices which are given explicitly as,
\begin{eqnarray}
\left(U_{(n)}\right)_{ab}=\, \delta_{a,b} 
\omega^{a}_{(n)} & \qquad{} \qquad{} &
\left(V_{(n)}\right)_{ab}= \delta^{(n)}_{a,b-1} \nonumber \\
\label{clock}
\end{eqnarray}
where $\omega_{(n)}=\exp(2\pi i/n)$ is an $n\,$'th root of unity, and 
$\delta^{(n)}$ denotes a modified Kronecker $\delta$ which is one if 
its two indices are equal modulo $n$ and is zero otherwise. 
An alternative way of specifying a vacuum state on ${\cal H}_{m}$ 
is by giving the expectation values of three independent gauge-invariant 
chiral operators,   
\begin{equation}
\begin{array}{ccc} 
\left\langle \frac{1}{N}{\rm Tr}\left[\Phi_{1}^{n}\right]\right\rangle 
= \alpha^{n}_{1} \qquad{} & 
\left\langle \frac{1}{N}{\rm Tr}\left[\Phi_{2}^{n}\right]
\right\rangle=\alpha^{n}_{2} 
\qquad{} &  \left\langle \frac{1}{N}{\rm Tr}\left[\Phi_{3}^{n}\right]
\right\rangle=\alpha^{n}_{3} \end{array}
\label{vevs}
\end{equation} 
\paragraph{}
The full Higgs branch ${\cal H}_{m}$ is the three dimensional 
complex orbifold ${\bf C}^{3}/{\bf Z}_{n}\times {\bf Z}_{n}$. 
The Higgs branch can be realized in string theory by placing $m$ 
coincident D3 branes at a point in the transverse space of a 
${\bf C}^{3}/{\bf Z}_{n}\times {\bf Z}_{n}$ singularity with one unit of 
discrete torsion. The effective action for the 
massless degrees of freedom on this branch is 
${\cal N}=4$ SUSY Yang-Mills with gauge group $U(m)$ and complex 
gauge coupling $n\tau$. We can also find vacua where the unbroken 
gauge symmetry is $U(1)^{m}$ by moving onto the Coulomb branch of the 
low-energy theory. This corresponds to separating the $m$ D3 branes in the 
string theory set-up.       
\paragraph{}
One way to characterise the different phases of a gauge theory is to 
probe the theory with external electric and magnetic charges. 
As usual, the possible 
electric charges 
are classified by the center of the gauge group \cite{TH}. 
The center of $U(mn)$ is $U(1)$ and the corresponding electric 
charge is denoted $q\in {\bf Z}$. Possible magnetic charges are classified by 
first homotopy class of the gauge group: 
$\pi_{1}\left(U(mn)\right)\simeq {\bf Z}$. We denote the corresponding 
magnetic charge $\tilde{q}\in{ \bf Z}$. If we move onto the Higgs branch, 
the vacuum condensate leads to the screening of external electric charges. 
As a $U(m)$ subgroup remains unbroken, only electric 
charges with $q=0$ mod $m$ are competely screened. Charges 
$q\neq 0$ mod $n$ will produce long-range Coulomb fields.  
The Higgs mechanism also leads to the 
confinement of external magnetic charges by the formation 
of chromomagnetic flux tubes. The unbroken $U(m)$ subgroup 
means that confinement is only partial. In particular those 
magnetic charges with $\tilde{q}=0$ mod $n$, produce long-range magnetic 
fields and remain unconfined.   
     
\subsection{The quantum theory}
\paragraph{}    
The quantum theory corresponding to the $\beta$-deformed superpotential 
(\ref{LSsup}) has several remarkable properties. Firstly it corresponds 
to an exactly marginal deformation of ${\cal N}=4$ SUSY Yang-Mills. 
More precisely, the theory has a critical surface 
in coupling constant space defined by $\kappa=
\kappa_{cr}[\tau,\beta]$ on which the all $\beta$-functions vanish and 
conformal invariance is unbroken. The critical surface 
includes the ${\cal N}=4$ line parametrized by the gauge coupling $\tau$ 
with $\beta=0$, $\kappa=1$. Thus we have a 
two-complex parameter family of ${\cal N}=1$ superconformal theories. 
\paragraph{}
As usual for ${\cal N}=1$ theories the exact vacuum structure is 
determined by the F-terms in the effective action which are holomorphic 
in the complex parameters $\beta$ and $\tau$. An exact solution for the 
holomorphic sector of the theory was obtained in \cite{DHK} and applied 
to the Higgs branch theory in \cite{D1}. 
One result is that the 
classical Higgs branch discussed in the previous section persists 
in the quantum theory for all values of the gauge coupling. 
The exact low energy gauge coupling on this branch is not renormalised and 
takes its classical value $n\tau$.           
\paragraph{}
As reviewed in \cite{D1}, the exact $SL(2,Z)$ duality of the ${\cal N}=4$ 
theory extends to the $\beta$-deformed theory. To be precise, the 
duality acts on a renormalized gauge coupling; 
\begin{equation}
\tau_{R}=\frac{4\pi i}{g^{2}_{R}}+\frac{\vartheta_{R}}{2\pi}
=\tau+ \frac{iN}{\pi}\log\kappa
\label{taur}
\end{equation}
and also on the deformation parameters as
\begin{equation}
\tau_{R}\rightarrow \frac{a\tau_{R}+b}{c\tau_{R}+d}  \qquad{} \qquad{} 
\beta \rightarrow 
\frac{\beta}{c\tau_{R}+d}  \qquad{} \qquad{} \kappa^{2}\sin\beta \rightarrow 
\frac{\kappa^{2}\sin\beta}{c\tau_{R}+d}
\label{sl2z}
\end{equation}
The algebraic renormalization of the coupling given in (\ref{taur}) plays no 
role in the following and we will ignore it from now on and supress the 
subscript on $\tau_{R}$. 
\paragraph{}
As usual the S-generator of $SL(2,Z)$, which acts as 
$\tau\rightarrow -1/\tau$ also interchanges electric and magnetic charges. 
Under this transformation, the 
Higgs phase vacuum 
where the gauge group is spontaneously broken down to a $U(m)$ 
subgroup is mapped to a confining phase with an 
unconfined $U(m)$ subgroup. Again we can characterise this phase 
by probing it with external charges. 
The S-duality transformation interchanges the integers $q$ and $\tilde{q}$ 
which characterize the possible electric and magnetic charges 
respectively. The magnetic condensate 
in the confining vacuum means that external magntic charges with 
$\tilde{q}=0$ mod $m$ are completely screened. Conversely, 
external electric charges are confined by the 
formation of chromoelectric flux tubes unless $q=0$ mod $n$. 
This phase is unusual because as it exhibits both electric confinement and 
spontaneously broken conformal invariance. 
Note that Higgs and Confining phases are genuinely different as expected 
in a theory containing only adjoint fields.     
\paragraph{}
Let us consider the action of the S-generator of 
$SL(2,Z)$ in the case $\vartheta=0$. The 
transformation relates the theory with parameters 
$g^{2}$ and $\beta=2\pi/n$ and chiral 
fields $\Phi_{i}$ to a dual theory with 
corresponding parameters $\tilde{g}^{2}=16\pi^{2}/g^{2}$ and 
$\tilde{\beta}=8\pi^{2}i/\tilde{g}^{2}n$ and chiral fields 
$\tilde{\Phi}_{i}$. The former theory has a Higgs branch. If we consider 
the Higgs branch vacuum with VEVs for 
$\Phi_{i}$ as given in (\ref{vacm},\ref{vevs}) 
above, then the S-dual vacuum 
has non-vanishing chiral VEVs, 
\begin{equation}
\begin{array}{ccc} 
\left\langle \frac{1}{N}{\rm Tr}\left[\tilde{\Phi}_{1}^{n}\right]\right\rangle 
= \tilde{\alpha}^{n}_{1} \qquad{} & 
\left\langle \frac{1}{N}{\rm Tr}\left[\tilde{\Phi}_{2}^{n}\right]
\right\rangle=\tilde{\alpha}^{n}_{2} 
\qquad{} &  \left\langle \frac{1}{N}{\rm Tr}\left[\tilde{\Phi}_{3}^{n}\right]
\right\rangle=\tilde{\alpha}^{n}_{3} \end{array}
\label{svevs}
\end{equation} 
The non-trivial modular weights of the chiral 
operators \cite{intril} appearing in (\ref{svevs}) imply that 
$\tilde{\alpha}_{i}=(\tilde{g}^{2}/4\pi)\alpha_{i}$ for $i=1,2,3$ \cite{D1}. 
\paragraph{} 
As explained in \cite{D1}, the scalar expectation values in (\ref{svevs})
do not correspond to any vacuum of the classical theory. At first sight 
the existence of this vacuum seems to lead to a contradiction for 
$\tilde{g}^{2}n<<1$ where one might expect the classical analysis to be 
valid. However, in this regime, the theory is weakly coupled only in the 
sense that the gauge coupling is small. 
In contrast, the deformation parameter $\tilde{\beta}$ is 
large and imaginary so that 
the Lagrangian includes exponentially large Yukawa couplings as well as 
quartic self-couplings of the adjoint scalars. 
Thus quantum corrections involving the adjoint scalars and their 
${\cal N}=1$ superpartners are not supressed and classical analysis is 
invalid. The confining phase vacua therefore lie on a quantum branch which is 
invisible classically.   
\section{Classical Deconstruction}
\paragraph{}
In \cite{D1}, the classical spectrum and effective action of the theory 
on the Higgs branch ${\cal H}_{m}$ was determined. As in \cite{D1}, we will 
consider the vacuum state on ${\cal H}_{m}$ specified by VEVs (\ref{vevs}) 
and, for simplicity, set $\alpha_{3}=0$. In this vacuum, the exact classical 
mass formula for each adjoint field is,    
\begin{equation}
M^{2}= 
4|\alpha_{1}|^{2}\sin^{2}\left(\frac{l_{1}\pi}{n}\right) + 
4|\alpha_{2}|^{2}\sin^{2}\left(\frac{l_{2}\pi}{n}\right)
\label{mass3} 
\end{equation}
with integers $l_{1}$, $l_{2}=1,2,\ldots, n$. 
\paragraph{}
Deconstruction starts from the observation that, for large-$n$, 
(\ref{mass3}) coincides with the spectrum of KK modes of 
a six-dimensional theory compactified to four dimensions on a torus. 
The integers $l_{1}$ and $l_{2}$ correspond to the quantized momenta 
around the two compact directions. At finite $n$, (\ref{mass3}) 
matches the truncated KK tower we would find if the extra dimensions were 
discretized on an $n\times n$ lattice. 
The appearance of extra dimensions is not 
limited to the spectrum but also extends to the classical action which 
can actually be rewritten as a six-dimensional gauge theory.       
In fact, the classical theory at fixed $n$ 
can be understood as a non-commutative $U(m)$ lattice gauge theory 
\cite{Sz, AMNS} defined on 
$R^{3,1}\times {\cal L}$, 
where ${\cal L}$ is an $n\times n$ lattice with 
periodic boundary conditions for all the fields. The lattice theory in 
question is a discretization of ${\cal N}=(1,1)$ supersymmetric 
non-commutative Yang-Mills theory with gauge group $U(m)$ compactified 
down to four dimensions on a torus. 
\paragraph{}
The parameters of the six dimensional theory 
can be expressed in terms of the four-dimensional parameters as follows. 
The lattice spacings of the two compact discrete dimensions are, 
\begin{eqnarray}
\varepsilon_{1}=\frac{1}{|\alpha_{1}|} & \qquad{} \qquad{} \qquad 
& \varepsilon_{2}=\frac{1}{|\alpha_{2}|} \nonumber \\
\label{table1}
\end{eqnarray}
The radii of the two-dimensional torus which they define are, 
\begin{eqnarray}
R_{1}=\frac{n}{2\pi |\alpha_{1}|} & \qquad{} \qquad{} \qquad{} 
& R_{2}=\frac{n}{2\pi|\alpha_{2}|} \nonumber \\
\label{table2}
\end{eqnarray}
The six-dimensional gauge coupling and non-commutivity parameter are given 
by, 
\begin{eqnarray}
G^{2}_{6}=\frac{g^{2}n}{|\alpha_{1}||\alpha_{2}|} & 
\qquad{} \qquad{} \qquad{} & 
\theta=\frac{n}{2\pi|\alpha_{1}||\alpha_{2}|}  \nonumber \\
\label{table3}
\end{eqnarray}
respectively. 
\paragraph{}
The derivation of these relations given in \cite{D1} was 
purely classical and, a priori they are only reliable at weak coupling 
$g^{2}N<<1$. The classical theory is 
characterised by the following heierarchy of length scales: 
$G_{6}<<\varepsilon_{i}<<\sqrt{\theta}<<R_{i}$. Provided we consider $n>>1$, 
these scales are well seperated and it makes sense to write down 
a six-dimensional continuum effective valid on length-scales much larger 
than the lattice spacing. This continuum effective action is 
precisely six-dimensional ${\cal N}=(1,1)$ supersymmetric gauge theory 
with gauge group $U(m)$ defined on $R^{3,1}\times T^{2}_{\Theta}$. 
Here $T^{2}_{\Theta}$ is the non-commutative torus with dimensionless 
non-commutativity parameter $\Theta=1/n$. The fact that 
$G_{6}<<\varepsilon_{i}$ shows that this effective 
continuum gauge theory is weakly coupled throughout its 
range of validity. 
\paragraph{}
A striking feature of the low-energy effective theory described 
above is that it has sixteen supercharges. In contrast the microscopic 
four-dimensional theory we started with only has four supercharges. 
The classical spectrum of W-bosons are BPS saturated with respect to the 
enlarged supersymmetry of the low-energy theory. 
The low-energy theory described above also has 
BPS saturated soliton strings which were studied in detail in \cite{D1}. 
These correspond to 
$U(m)$ Yang-Mills instantons on 
$R^{2}\times T^{2}_{\Theta}$ \cite{Nek} 
embedded as static field configurations 
in six dimensions. The lightest strings, 
corresponding to instantons of topological charge one, have tension,        
\begin{equation}
T=\frac{8\pi^{2}}{G_{6}^{2}}=\frac{8\pi^{2}|\alpha_{1}||\alpha_{2}|}{g^{2}n}
\label{tension}
\end{equation}
In terms of the underlying four-dimensional $U(N)$ gauge theory these 
strings are precisely the expected chromomagnetic flux tubes 
which confine external magnetic charges. In particular, an instanton 
string of topological charge $k$ can end on and external magntic charge 
$\tilde{q}=k$ mod $n$.    

\section{A First Look at the Continuum Limit}
\paragraph{}
Given any lattice theory it is natural to question whether we can find an 
interacting continuum limit. In the present context this means a limit in 
which the lattice spacing goes to zero, 
while the six-dimensional gauge coupling is held fixed. One interesting limit 
discussed in \cite{D1} is\footnote{This is the limit 
discussed in Section 9.2 of \cite{D1}. 
A different continuum limit which yields a non-commutative theory 
was also discussed in \cite{D1} but we will not consider it here.}, 
\paragraph{}
{\bf Limit I:} We consider the $U(mn)$ theory with $\beta=2\pi/n$ 
in the Higgs branch vacuum specified in (\ref{vevs}) above and take the limit 
$n\rightarrow \infty$, $g^{2} \rightarrow \infty$ 
while holding $m$ and $g^{2}/n$ fixed  We also scale the VEVs as 
\begin{eqnarray}
|\alpha_{1}| \sim n\rightarrow \infty & \qquad \qquad & 
|\alpha_{2}| \sim n\rightarrow \infty \nonumber \\
\label{scalingsI}
\end{eqnarray}
\paragraph{}
Using the results (\ref{table1},\ref{table2},\ref{table3}) for the 
six-dimensional parameters, we find that the lattice spacings 
$\varepsilon_{i}$ and the non-commutativity parameter $\theta$ go to zero, 
while the six-dimensional gauge coupling $G_{6}$ and the radii of 
compactification $R_{i}$ remain fixed. Naively this indicates that we 
end up with  a commutative continuum theory defined on $R^{3,1}\times T^{2}$. 
Of course Limit I, as defined above, is a strong coupling limit and 
we should immediately question whether it is legitimate to extrapolate 
the classical formulae (\ref{table1},\ref{table2},\ref{table3}) which were 
derived assuming weak coupling. We will postpone this 
discussion momentarily and take Limit I at face value as a 
candidate continuum limit.  
\paragraph{}
The next question is what continuum theory could possibly arise in this 
limit. The naive answer based purely on our classical analysis 
is that we find a conventional $U(m)$ gauge theory with ${\cal N}=(1,1)$ 
supersymmetry in six dimensions. This cannot be the whole story as such 
a theory is certainly non-renormalisable and requires a consistent UV 
completion to make sense. In fact there is only one candidate theory which 
provides such a completion without also coupling the low-energy theory 
to gravity. This is the Little String Theory discussed in Section 
2 above. The low-energy gauge group of LST is $SU(m)$ 
rather than $U(m)$ so we also 
need to include an additional free $U(1)$ vector multiplet of 
${\cal N}=(1,1)$ SUSY. The simplest deconstruction conjecture is therefore, 
\paragraph{}
{\bf Conjecture:} When we take Limit I, the 
theory on the Higgs branch with $\beta=2\pi/n$ 
becomes Type IIB LST on $R^{3,1}\times T^{2}$ plus an additional 
decoupled $U(1)$ vector multiplet. The mass scale of LST is given by, 
\begin{equation}
M_{s}=\sqrt{\frac{|\alpha_{1}||\alpha_{2}|}{g^{2}n}}
\label{msconf0}
\end{equation}              
which remains fixed in Limit I, as do the radii of compactification 
$R_{1}$ and $R_{2}$ given in (\ref{table2}) above.    
\paragraph{}
At this stage, the motivation for the conjecture depends 
on our extrapolation of the classical formulae 
(\ref{table1},\ref{table2},\ref{table3}) to strong coupling. 
As the classical low-energy effective action has sixteen supercharges 
it is tempting to try to invoke non-renormalisation theorems to justify 
this extrapolation. In particular, the spectrum of W-bosons 
which represent the Kaluza-Klein modes of the six-dimensional 
effective theory are BPS saturated with respect to this enlarged 
supersymmetry. The mass spectrum of these states 
dictates both the radii of compactification and the lattice spacing. 
As the masses of BPS states are protected from quantum corrections 
in a theory with sixteen supercharges, we might hope to infer that the 
classical formulae (\ref{table1},\ref{table2}) are exact. 
\paragraph{}
The argument given above is far from convincing. Although the 
low-energy effective action 
has sixteen supercharges, it is obtained by integrating out massive degrees 
of freedom starting from the full action of the $\beta$-deformed theory. 
As the latter has only ${\cal N}=1$ supersymmetry and thus only 
the F-terms in the 
effective action are protected. This includes the low-energy gauge 
coupling, but not the radius of compactification or the effective 
lattice spacing which certainly depend on D-terms.   
\paragraph{}
To make a more convincing version of the above
non-renormalisation argument we can focus on a limit in which the theory 
recovers sixteen supercharges at all length scales. 
As the Higgs branch theory has deformation parameter $\beta=2\pi/n$, the 
microscopic Lagrangian goes over to that of the ${\cal N}=4$ theory in a 
limit $n\rightarrow \infty$. As the masses of 
BPS states are protected in the ${\cal N}=4$ theory, one can argue 
that any quantum corrections to the classical mass formula are 
supressed by powers of $1/n$. The resulting classical formulae 
(\ref{table1},\ref{table2}) should then become exact in the 
large-$n$ limit. Unfortunately, even this argument will not help 
us understand the proposed continuum limit. In particular, Limit I    
is a  simultaneous large-$n$ and strong coupling limit. Although corrections 
to the classical formulae which go like $1/n$ may be supressed, 
those which go like $g^{2}/n$ are not and may be important. 
\paragraph{}
The points raised above indicate that the existence and nature of the 
continuum limit depends on the detailed dynamics of the strongly coupled 
gauge theory. The main aim of this paper is to study the continuum limit  
in a controlled way. As suggested in \cite{D1} the first step is to perform 
an S-duality transformation a reinterpret the strongly coupled Higgs 
branch theory in terms of the S-dual confining phase. The S-dual of 
Limit I is,   
\paragraph{}
{\bf Limit \~I:} We consider the $U(mn)$ theory with gauge coupling coupling 
$\tilde{g}^{2}$, zero vacuum angle $\tilde{\vartheta}=0$ 
and deformation parameter $\tilde{\beta}=8\pi^{2} i/\tilde{g}^{2}n$. 
We focus on the theory in 
the confining phase vacuum specified by non-zero VEVs (\ref{svevs}).  
We take the limit $n\rightarrow \infty$, $\tilde{g}^{2} \rightarrow 0$ 
while holding $m$ and $\tilde{g}^{2}n$ (and thus $\tilde{\beta}$) 
fixed. We also hold the VEVs  $|\tilde{\alpha}_{1}|$ and 
$|\tilde{\alpha}_{2}|$ fixed.   
\paragraph{}
Thus we see that S-duality has three notable effects. First, 
because of the non-trivial modular transformation of the scalar VEVs, the 
new limit is one where the dimensionful parameters 
$\tilde{\alpha}_{i}$ labelling the vacuum are held fixed. 
Similarly, 
although the deformation parameter $\beta=2\pi/n$ goes to zero in Limit I, 
the dual parameter $\tilde{\beta}=8\pi^{2} i/\tilde{g}^{2}n$ remains 
fixed in the S-dual Limit \~I. This already illustrates one 
of the points made above: at fixed $\tilde{g}^{2}n$ the 
S-dual theory only has ${\cal N}=1$ 
supersymmetry in the UV. Thus the naive argument  
that ${\cal N}=4$ supersymmetry is automatically recovered in Limit I  
simply because $\beta$ goes to zero is wrong!
Finally we see that S-duality maps 
a strong coupling limit of the Higgs phase theory to a 't Hooft 
limit of a confining phase theory. As the confinement 
is only partial, standard large-$N$ scaling argument do not 
immediately apply. In Section 9, we will reconsider this issue in detail 
in the context of deconstructing double-scaled LST.    
\paragraph{} 
The S-duality transformation also raises a puzzling issue about the 
effective lattice spacing in the confining phase. As in the 
Higgs phase, the scale invariance of the UV theory is broken only 
by the non-zero scalar vacuum expectation values (\ref{svevs}). 
Thus we would expect the theory exhibit approximate conformal invariance 
above the energy scale, 
\begin{equation}          
E \sim {\rm max}[\tilde{\alpha}_{1},\tilde{\alpha}_{2}]
\label{estimate}
\end{equation}
On the other hand, from the point of view of deconstruction, we 
would expect the scale where four-dimensional conformal invariance is 
recovered to be set by the smallest lattice spacing 
$\varepsilon={\rm min}[\varepsilon_{1},\varepsilon_{2}]$. 
Rewriting the classical formula (\ref{table1}) in terms of the 
confining phase variables, this corresponds to an energy scale,    
\begin{equation}          
E \sim \varepsilon^{-1}=
\left(\frac{4\pi}{\tilde{g}^{2}}\right)
{\rm max}[\tilde{\alpha}_{1},\tilde{\alpha}_{2}]
\label{estimate2}
\end{equation}
\paragraph{}
This discrepancy has an obvious origin. On the Higgs branch the 
lattice spacing is set by the masses of the heaviest W-boson. 
Assuming the classical 
mass formula is valid, we expect this state to become very massive 
in the strong-coupling limit. 
The corresponding lattice spacing would then go to zero. 
From the point of view of the confining phase this state 
is a magnetic monopole which is very heavy for small values of the dual 
coupling. On the other hand we might expect that, as in the underlying 
${\cal N}=4$ theory, the scale at which conformal invariance is broken 
is actually set by the elementary electric degrees of freedom 
rather than the magnetic ones. These two possibilities  
lead to the two different formulae (\ref{estimate}) and 
(\ref{estimate2}).  
\paragraph{}    
In the following we will clarify the issues raised above by explicit 
calculation. Specifically, if we choose $\tilde{g}^{2}n>>1$, 
then the proposed continuum limit 
lies in a regime where we can hope to study it directly using the 
AdS/CFT correspondence. As the deformation parameter 
$\tilde{\beta}=8\pi^{2} i/\tilde{g}^{2}n$ is then small, we can think about 
the conformal theory as a small deformation of the ${\cal N}=4$ theory. 
On the other hand, as the 't Hooft coupling is large, 
the latter is well described by IIB 
supergravity on $AdS_{5}\times S^{5}$. The deformation can be 
incorporated by introducing a source for the corresponding 
supergravity field on the boundary of $AdS_{5}$. The main aim of the 
next two sections will be to construct an explict AdS dual for the 
$\beta$-deformed theory on its confining branch.   

\section{String Theory Dual of the Higgs Branch}
\paragraph{}
In this Section we will construct the string dual of the $\beta$-deformed 
theory. First we review the embedding of the weakly-coupled Higgs branch  
in IIB string theory discussed in \cite{D1}. We then move on to consider 
the AdS dual of the strongly coupled Higgs branch theory. Finally, in the next Section, we will use IIB 
S-duality to find the corresponding dual for the confining phase theory.  

\subsection{Weak Coupling Analysis}
\paragraph{}
It is straightforward to embed the 
$\beta$-deformed theory in string theory \cite{D1}
starting from the standard realization of ${\cal N}=4$
SUSY Yang-Mills theory with gauge group $U(N)$ on the world-volume of a 
stack of $N$ D3 branes in Type IIB string theory. As usual, 
the string coupling is related to the four-dimensional gauge coupling as 
$g_{s}=g^{2}/4\pi$. For convenience we will set the 
field theory vacuum angle $\vartheta$ (which is 
proportional to the background value of the RR scalar $C_{(0)}$) to 
zero in the following. In this subsection, we will 
initially work at small 't Hooft coupling $g^{2}N=4\pi g_{s}N<<1$, 
so that the gauge theory on the branes is weakly coupled.    
\paragraph{}
The $\beta$-deformation is 
introduced by turning on a particular background for the 
complex threeform field-strength,
\begin{equation}
G_{(3)}=F_{(3)}-\tau H_{(3)}
\label{g32}
\end{equation}
which will be described in more detail in subsection 6.2 below
\footnote{At the linearized level, the relation between background 
supergravity fields and deformations of the ${\cal N}=4$ theory on the 
brane in weakly coupled string theory is essentially the same as the 
standard dictionary between SUGRA fields and chiral primary operators 
provided by the AdS/CFT correspondence \cite{tvr}.}. 
Here $F_{(3)}$ and $H_{(3)}$ are the RR and NS three-form field 
strengths respectively.
For small deformations the 
field strength is proportional to the deformation parameter $\beta$ with a 
real constant of proportionality. 
Thus the theory with $\beta=2\pi/n$ 
corresponds to a non-zero field strength for $F_{(3)}$ of order $1/n$. 
For large-$n$ the deformation can be treated as a small perturbation   
and the back-reaction of this field strength on the geometry can be 
neglected.
\paragraph{}
At weak coupling, the Higgs branch which appears for $\beta=2\pi/n$ 
corresponds to configurations 
where the D3 branes polarize into $m$ D5 branes wrapped on 
a torus in the transverse $R^{6}$ \cite{D1}. We define convenient 
complex combinations of the coordinates, 
$x_{m}$ with $m=4,\ldots 9$, in the six transverse dimensions, 
\begin{equation}
\begin{array}{ccc}
z_{1}= x_{4}+ix_{7}=
\rho_{1}e^{i\psi_{1}} & z_{2}=x_{5}+ix_{8}=\rho_{2}e^{i\psi_{2}} & 
z_{3}=x_{6}+ix_{9}=\rho_{3}e^{i\psi_{3}} 
\end{array}  
\label{complex2}
\end{equation}
In the vacuum (\ref{vacm}) with $\alpha_{3}=0$, 
the D5 branes are wrapped on a rectangular torus 
$T^{2}(r_{1},r_{2})$ defined in terms of 
these coordinates by the 
equations, 
\begin{equation}
\begin{array}{ccc}
\rho_{1}=r_{1}=|\alpha_{1}|(2\pi \alpha') \qquad{} & 
\rho_{2}=r_{2}=|\alpha_{2}|(2\pi \alpha') \qquad{} & \rho_{3}=0 
\end{array}
\label{tr1}
\end{equation} 
This configuration is energetically stable because of the 
non-zero background flux.        
\paragraph{}
The configuration of wrapped D5's also carries a total of $N$ units of 
D3 brane charge which can be realized as a constant 
non-zero background for the gauge-invariant combination, 
\begin{equation}
{\cal F}= F- \frac{1}{2\pi\alpha'}B_{\rm NS}
\label{gind2}
\end{equation}
where $F$ is the world-volume gauge field-strength in the central $U(1)$ of 
$U(m)$ and $B_{\rm NS}$ is the 
Neveu-Schwarz two-form potential. The presence of $N$ D3 branes then 
implies, 
\begin{equation}
\int_{T^{2}}\, {\cal F}= 2\pi n
\label{quantum4}
\end{equation}
\paragraph{}
We introduce flat coordinates $\chi_{1}$ and $\chi_{2}$ on the torus 
with $0\leq \chi_{1}\leq 2\pi r_{1}$ and $0\leq \chi_{2}\leq 2\pi r_{2}$. 
In this basis the metric on the torus is the flat one 
$g_{ab}=\delta_{ab}$. Choosing a gauge $F=0$, the two-form field 
can be written as, 
 \begin{equation}
B_{\rm NS}=\frac{n}{2\pi r_{1}r_{2}}\, (2\pi\alpha')\, d\chi_{1}\wedge d\chi_{2} = 
\frac{n}{2\pi |\alpha_{1}||\alpha_{2}|}\, \frac{1}{(2\pi \alpha ')} 
\, d\chi_{1}\wedge d\chi_{2}                                                   
\label{bns4}
\end{equation}
\paragraph{}
As usual the presence of $B_{\rm NS}$ introduces non-commutivity in the 
$T^{2}$ component of the $D5$ world volume. Standard arguments imply that the 
low-energy gauge theory on the $D5$ brane worldvolume six-dimensional 
non-commutative Yang-Mills theory with sixteen supercharges. Indeed one can 
check \cite{AF1,D1} that this coincides precisely with the classical 
effective theory described in Section 4, with parameters exactly as 
given in (\ref{table2},\ref{table3}). 

\subsection{Strong Coupling Analysis}
\paragraph{}
In the previous subsection we reviewed the IIB brane configuration 
which realizes the weakly-coupled Higgs branch for $n>>1$ and $g^{2}N<<1$. 
In this subsection we will 
use the ideas of \cite{PS} to follow this configuration to the regime of 
strong 't Hooft coupling $g^{2}N>>1$. As before we have $N=mn$ and choose 
$n>>1$. In this case the Higgs branch deformation parameter $\beta=2\pi/n$ 
is small and we will work perturbatively in $|\beta|$.  
We start by considering the undeformed ${\cal N}=4$ theory with 
$\beta=0$. At large 't Hooft coupling this is dual to 
the near horizon geometry of $N$ D3 branes. The corresponding 
IIB supergravity background has the general form:    
\begin{eqnarray}
ds^{2} \,\,  & = & {\cal Z}^{-\frac{1}{2}}\eta_{\mu\nu}
dx^{\mu}dx^{\nu} +     {\cal Z}^{+\frac{1}{2}}dx_{m}dx_{m} \nonumber \\
F_{5} & = & d\chi_{4}+ \star d\chi_{4} \nonumber \\
\chi_{4} &= & 
\frac{1}{g_{s}{\cal Z}}\,\, dx_{0}\wedge dx_{1}\wedge dx_{2}\wedge 
dx_{3} \nonumber \\ 
e^{\phi}=g_{s} & \qquad{} \qquad{} & C_{(0)}=
\frac{\vartheta}{2\pi} \nonumber \\ 
\label{coulomb}
\end{eqnarray}  
The coordinate indices $x_{\mu}$, with $\mu=0,1,2,3$, are the 
four spacetime coordinates parallel to the branes and $x_{m}$, with 
$m=4,5,\ldots,9$ denote the six transverse directions. In the above 
$\chi_{4}$ is the RR four-form potential and $F_{5}$ is the corresponding 
self-dual five-form field strength. Here ${\cal Z}$ denotes a 
harmonic function of the radial coordinate $r=\sqrt{x_{m}x_{m}}$.
\paragraph{} 
The ${\cal N}=4$ theory with gauge group $SU(N)$ at its conformal point is 
dual to the near horizon geometry of $N$ coincident D3 branes. This 
corresponds to the choice, ${\cal Z}_{\rm UV}={\cal R}^{4}/r^{4}$ 
in (\ref{coulomb}) 
with ${\cal R}^{4}=4\pi g_{s}N \alpha'^{2}$. The resulting 
geometry is precisely $AdS_{5}\times S^{5}$ with radius ${\cal R}$ for both 
factors. Gauge theory and string theory parameters are 
related according to, 
\begin{eqnarray}
g^{2}=4\pi g_{s} & \qquad{} \qquad{} \qquad 
& g^{2}N=\frac{{\cal R}^4}{\alpha'^{2}}\nonumber \\
\label{ident4}
\end{eqnarray}
and the classical supergravity approximation is valid for our chosen regime 
$g^{2}<<1$, $g^{2}N>>1$.  
\paragraph{} 
As in the previous section, we are interested is a configuration 
where the $N$ D3 branes are uniformly distributed on the torus 
$T^{2}(r_{1},r_{2})$ defined in (\ref{tr1}). In $\beta=0$ case, this 
corresponds to a 
particular point on the Coulomb branch of the ${\cal N}=4$ theory.    
The corresponding near-horizon geometry takes the 
form (\ref{coulomb}) with the harmonic function ${\cal Z}(r)$ given as 
\cite{coulomb} 
\begin{eqnarray}
{\cal Z} & = & \frac{{\cal R}^{4}}{4\pi^{2}} \int_{0}^{2\pi} \, 
d\psi_{1} \, \int_{0}^{2\pi}\, d\psi_{2}\, \frac{1}{D(\psi_{1},\psi_{2})^{2}} 
\nonumber \\
\label{bigint}
\end{eqnarray}    
with, 
\begin{equation}
D(\psi_{1},\psi_{2})=\rho_{1}^{2}-2\rho_{1}r_{1}\cos\psi_{1}+ r_{1}^{2}     
+\rho_{2}^{2}-2\rho_{2}r_{2}\cos\psi_{1}+r_{2}^{2}+ \rho_{3}^{2}
\label{dpsi}
\end{equation}
This geometry provides a dual description of the Coulomb branch theory 
at large 't Hooft coupling. 
Note that ${\cal Z}\rightarrow {\cal Z}_{\rm UV}= {\cal R}^{4}/r^{4}$ 
as $r\rightarrow \infty$ and so the geometry asymptotes to 
$AdS_{5}\times S^{5}$ for large $r$. This 
corresponds to the RG flow of the dual theory to the ${\cal N}=4$ 
superconformal fixed point in the UV.
\paragraph{}
The geometry also simplifies in the region very close to the branes, which 
corresponds to, 
\begin{equation}
\begin{array}{ccc} \rho_{1}-r_{1}<< r_{1} & \rho_{2}-r_{2} << r_{2} & 
\rho_{3} << r_{1} {\rm ,}\, r_{2}
\end{array}
\label{nearshell}
\end{equation}
In this region the integral (\ref{bigint}) is dominated by the neighbourhood 
of the singular point $\psi_{1}=\psi_{2}=0$ and the corresponding 
warp-factor reduces to, 
\begin{eqnarray}
{\cal Z}_{\rm IR} & = & \frac{{\cal R}^{4}}{4\pi r_{1}r_{2}v^{2}} \\
\label{zir}
\end{eqnarray}
where, 
\begin{equation}
v=\sqrt{(\rho_{1}-r_{1})^{2}+(\rho_{2}-r_{2})^{2}+ \rho_{3}^{2}}
\label{d}
\end{equation} 
is the distance from the toroidal shell of D3 branes.
\paragraph{}
As mentioned above, introducing the $\beta$-deformation corresponds to 
turning on a non-zero background for the complex three-form field strength 
$G_{(3)}$ proportional to the deformation parameter $\beta$. 
We will be mainly interested in the case of $|\beta|<<1$ which corresponds 
to introducing small field strength on the boundary of $AdS_{5}$. 
In the conformally invariant vacuum, this leads to a small deformation of 
the $AdS_{5}\times S^{5}$ geometry 
which can be constructed order by order in $|\beta|$. This program was 
carried out explicitly to second non-trivial order in \cite{AKY}. As the 
deformed theory is ${\cal N}=1$ superconformal invariant one finds 
a dual of the form $AdS_{5}\times \tilde{S}^{5}$. Here $\tilde{S}^{5}$ 
denotes a small deformation of $S^{5}$. The resulting perturbation away 
from the the round metric on $S^{5}$ reflects the 
fact that the $SU(4)$ R-symmetry of the ${\cal N}=4$ theory is broken to 
$U(1)^{3}$ for non-zero $\beta$. 
\paragraph{}
In the present case we will only need to work to linear order in the 
deformation. 
The resulting three-form flux can be deduced from the analysis of Gra\~{n}a 
and Polchinski \cite{GP}. 
In fact we will only need the 
``self-dual'' component of the field-strength,  
\begin{equation}
G_{(3)}^{+}=G_{(3)}+i \star_{6}G_{(3)} 
\label{sdg}
\end{equation}
where $\star_{6}$ denotes the Hodge dual in the transverse $R^{6}$ 
parameterized by coordinates $x_{m}$ appearing in (\ref{coulomb}) 
with the flat metric 
$\delta_{mn}$. At linear order the resulting flux can be written as, 
$G_{(3)}^{+}=(i\rho\beta/3g_{s}){\cal Z}(r)d\omega_{2}$ where $\rho$ is a real 
constant we have not determined and $\omega_{2}$ 
is a two-form which is conveniently written in terms of the complex 
coordinates introduced in (\ref{complex2}) above. Explicitly we have, 
\begin{equation} 
\omega_{2}=z_{1}z_{2}d\bar{z}_{1}\wedge d\bar{z}_{2} 
+z_{2}z_{3}d\bar{z}_{2}\wedge d\bar{z}_{3} +
z_{3}z_{1}d\bar{z}_{3}\wedge d\bar{z}_{1}  
\label{2form}
\end{equation}
This result holds in any asymptotically AdS background of the form 
(\ref{coulomb}) and we will apply it with the warp factor ${\cal Z}$ 
chosen as in (\ref{bigint}). The criterion for a 
small deformation of the $AdS_{5}\times S^{5}$ geometry is $|\beta|<<1$, 
which matches the condition for a small deformation of the ${\cal N}=4$ 
theory on the boundary. In the following, we will take account of 
this non-zero three-form background and its effect on probe branes 
in the bulk. However, as we are working to linear order in $|\beta|$, we 
will not need to consider the back-reaction on the geometry explicitly which 
appears at order $|\beta|^{2}$.   
\paragraph{}
We are now ready to construct the AdS dual of the Higgs branch theory. 
As at weak coupling our starting point is $m$ 
D5 branes wrapped on the torus $T^{2}[r_{1},r_{2}]$. 
The torus also carries $N$ units of D3 brane charge. 
As above, the backreaction of the D3 branes leads to a near-horizon 
geometry of the form (\ref{coulomb}) 
with the non-trivial warp factor (\ref{bigint}).
In a complete treatment of the full system we would need to 
account for the backreaction of both the D3 branes and the 
wrapped D5 branes. This would be a very hard problem. Fortunately, 
as in \cite{PS}, we may choose work in a regime of parameters where 
the problem simplifies. 
\paragraph{}
To illustrate this we consider D5 branes embedded in the 
$AdS_{5}\times S^{5}$ geometry corresponding to the large-$r$ warp factor 
${\cal Z}_{\rm UV}$ defined above. In this metric, 
the torus $T^{2}[r_{1},r_{2}]$, defined in (\ref{tr1}), is located at 
fixed radial position $r=\bar{r}=\sqrt{r_{1}^{2}+r_{2}^{2}}$ in $AdS_{5}$. 
Our probe configuration becomes $m$ D5 branes 
located at $r=\bar{r}$ and wrapped on a torus $T^{2}\subset S^{5}$.  
The area of the torus is of order 
${\cal R}^{2}=\sqrt{4\pi g_{s}N}\alpha'$. The average density 
of D3 brane charge is therefore $\sigma_{3}\sim n/{\cal R}^{2}$. 
This gives, 
\begin{equation}        
\sigma_{3} \sim \sqrt{\frac{n}{g_{s}m}}\, \frac{1}{\alpha'}
\label{ratio}
\end{equation}
It follows that provided $g_{s}/n<<m$, the energy density of D3 branes 
is large in string units and dominates that of D5 branes. 
This condition is trivially satisfied in our chosen region of parameter space 
\paragraph{}
The above argument suggests that, for $g_{s}/n<<m$, it is legitimate to treat 
the D5 branes as probe branes wrapped on a torus $T^{2}\subset S^{5}$ 
in the Coulomb branch geometry (\ref{coulomb},\ref{bigint}). 
The first obvious test of this reasoning is to check 
that such a configuration is stable. In the dual field theory, the 
Higgs branch only exists for special values of the deformation parameter. 
The brane configuration should therefore be stable only for the 
corresponding values of the complex three-form flux.
As the complex moduli $\alpha_{1}$ and $\alpha_{2}$ are flat directions of 
the potential for $\beta=2\pi/n$, the resulting configuration should then be 
stable for all values of the radii $r_{1}$ and $r_{2}$.  
In Appendix A we perform a probe calculation using the Dirac-Born-Infeld 
action of the D5 brane and verify these expected 
properties. 
\paragraph{}
In conclusion, the AdS dual of the Higgs branch 
consists of $m$ wrapped D5 branes in the Coulomb branch geometry. 
In addition a small background flux for the RR threeform is present 
which has the role of stabilizing the the wrapped branes. As discussed above 
the backreaction of the D5 branes on the geometry can be consistently 
neglected almost everywhere. As in \cite{PS}, the only exception to this 
is a thin layer close to the branes where the expansion of the metric in 
two directions transverse to the D3s and parallel to the D5s effectively 
dilutes the D3 brane charge. In this region 
the fivebrane charge effectively dominates. The SUGRA solution in this 
regime should match onto the near-horizon 
geometry of $m$ flat D5 branes with world-volume $B_{\rm NS}$. 
\paragraph{}
Following Polchinski and Strassler, our approximate solution is 
obtained by patching together the solution in 
the two regions described above.   Thus we 
compare the Coulomb branch solution (\ref{coulomb}) for 
${\cal Z}={\cal Z}_{\rm IR}$ given above with the UV behaviour of the 
D5 brane background (\ref{d5soln2}). In the limit $u>>a^{-1}$ we have 
$\bar{h}(u)\simeq 1/a^{2}u^{2}$ and (\ref{d5soln2}) can be written as, 
\begin{eqnarray}
ds^{2} \,\,  & = & \alpha' \left(\frac{s}{\sqrt{bg_{s}m}}\right)
\eta_{\mu\nu}dy^{\mu}dy^{\nu}+ 
\frac{1}{\alpha'}\left(\frac{\sqrt{bg_{s}m}}{s}\right)dy_{m}dy_{m} 
\nonumber \\
B_{\rm NS} \,\, & = & \frac{b}{\alpha'} \,\,dy_{4}
\wedge dy_{5} \nonumber \\ 
\exp(2\phi) \,\, & = & \, \,g_{s}^{2} 
\nonumber \\
\chi_{4} \,\, & = & \frac{1}{g_{s}} \left(\frac{s^{2}}
{bg_{s}m}\right)\,\,  
dy_{0}\wedge dy_{1}\wedge dy_{2}\wedge dy_{3} \nonumber \\
\label{d5soln3}
\end{eqnarray} 
Remarkably we find that these two metrics match exactly if we identify 
the coordinates as $y_{\mu}=x_{\mu}$ for $\mu=0,1,2,3$ and 
$y_{m}=M_{mn}x_{n}+N_{mn}$, where $M$ is an $SO(6)$ rotation matrix and 
$N_{mn}$ is another constant matrix, 
for $m,n=4,5,\ldots,9$. This implies that the radial 
coordinate $s=\sqrt{y_{6}^{2}+\ldots +y_{9}^{2}}$ 
is identified with the distance $v$ from the branes. 
The dilaton and five-form solutions  
also match and the constant $B_{\rm NS}$ field in (\ref{d5soln3}) can be 
removed with a gauge transformation. Comparing the coefficients appearing in 
the metric we find that,  
\begin{equation}
b=\frac{\alpha'^{2}n}{r_{1}r_{2}}=\frac{n}
{4\pi^{2}|\alpha_{1}||\alpha_{1}|}
\label{match} 
\end{equation}
\paragraph{}
The resulting solution can be written as, 
\begin{eqnarray}
ds^{2} \,\,  & = & {\cal Z}_{\rm IR}^{-\frac{1}{2}}\eta_{\mu\nu}
dx^{\mu}dx^{\nu} + {\cal Z}_{\rm IR}^{+\frac{1}{2}}
\left(\frac{v^{2}}{v_{cr}^{2}+v^{2}}\right)[dx_{4}^{2}+dx_{5}^{2}] 
+{\cal Z}_{\rm IR}^{+\frac{1}{2}}[dv^{2}+v^{2}d\Omega_{3}^{2}] \nonumber \\
B_{\rm NS} \,\, & = &\frac{b}{\alpha'}
\left(\frac{v^{2}}{v_{cr}^{2}+v^{2}}\right)  \,\,dx_{4}
\wedge dx_{5} \nonumber \\ 
e^{2\phi} & = & g^{2}_{s}\left(\frac{v^{2}}{v_{cr}^{2}+v^{2}}\right)  
\nonumber \\ \chi_{4} &= & 
\frac{1}{g_{s}{\cal Z}_{\rm IR}}\,\, dx_{0}\wedge dx_{1}\wedge dx_{2}\wedge 
dx_{3} \nonumber \\ 
\label{coulomb2}
\end{eqnarray}  
where ${\cal Z}_{\rm IR}$ is given in (\ref{zir}) and the crossover occurs at  
\begin{equation}
v_{cr}=\sqrt{\frac{g_{s}m r_{1}r_{2}}{n}}
\label{vcr}
\end{equation}
The proceedure of patching the two solutions together makes sense provided 
we can find a range of values where the two asymptotic forms used are 
simultaneously valid. This is equivalent to the condition $v_{cr}<<r_{1}$, 
$r_{2}$ which holds provided $g_{s}m/n<<1$. The latter condition is trivially 
satisfied in the limit of interest (large-$n$ with $g_{s}<<1$). 
Finally, the toroidal compactification of the worldvolume theory is 
implimented by a periodic identification of the 
coordinates in the $x_{4}$ and $x_{5}$ directions, 
\begin{eqnarray}
x_{4}\sim x_{4}+2\pi r_{1}   & \qquad{} \qquad{} & 
x_{5}\sim x_{5}+2\pi r_{2} 
\label{ident45}
\end{eqnarray}
\paragraph{}
Thus we find a geometry which interpolates between $AdS_{5}\times S^{5}$ 
far away from the wrapped branes and a geometry which coincides with the 
holographic dual of the decoupled five-brane theory 
${\cal T}[M_{s},g_{s}]$. Using the matching (\ref{match}) we can identify 
the parameters of the five-brane theory in terms of those of the 
four-dimensional gauge theory. As always we have $g_{s}=g^{2}/4\pi$ and 
in addition we find, 
\begin{equation}
M_{s}=\sqrt{\frac{|\alpha_{1}||\alpha_{2}|}{g^{2}n}}
\label{mshiggs}
\end{equation}
at low energies the fivebrane theory ${\cal T}[M_{s},g_{s}]$ reduces to 
a non-commutative six dimensional gauge theory with $G_{6}=1/M_{s}$ and 
$\theta=1 /8\pi^{2}g_{s}M_{s}^{2}$. 
This precisely reproduces the values 
of these parameters (\ref{table3}) computed in weakly coupled field theory. 
As at weak coupling, the compactification radii of the worldvolume 
theory should be measured in the open string metric. As in \cite{D1}, 
converting 
the closed string radii $r_{1}$ and $r_{2}$ to the corresponding 
lengths in the open string metric reproduces the classical formula 
(\ref{table2}) for the radii $R_{1}$ and $R_{2}$.     
\paragraph{}
We can now try to interpret our solution as 
describing an RG flow between four-dimensional superconformal 
field theory in the UV and six dimensional non-commutative gauge theory in 
the IR. In the UV region $r>>\bar{r}$ where conformal invariance is 
restored, the UV/IR connection provides a relation between the 
radial dependence of the supergravity solution and RG flow in the dual 
field theory. As the deviations from the $AdS_{5}$ geometry become 
significant at $r\sim \bar{r}$ we identify the corresponding energy scale, 
\begin{equation}
\Lambda=\frac{\bar{r}}{\sqrt{g_{s}N}\alpha'} \, \sim \, 
\frac{1}{\sqrt{g^{2}N}}\,
{\rm max}\left[|\alpha_{1}|,|\alpha_{2}|\right]         
\end{equation}
with the scale in the dual field theory below which the breaking of conformal 
symmetry becomes significant. Above this mass-scale, the theory looks like 
a four-dimensional conformal field theory with no trace of six dimensional 
behaviour. Equivalently, 
this scale is the effective momentum-space cutoff on the 
low-energy six-dimensional theory. In the language of deconstruction 
it is natural to identify this with the inverse of the 
smaller lattice-spacing 
$\varepsilon={\rm min}[\varepsilon_{1},\varepsilon_{2}]$. 
At weak coupling the corresponding scale is given   
by (\ref{table1}) as, 
\begin{equation}
\Lambda_{\rm WC}= \frac{1}{\varepsilon}=
{\rm max}\left[|\alpha_{1}|,|\alpha_{2}|\right]     
\label{lamwc}
\end{equation} 
Thus already we see a significant (but quite familiar) 
discrepancy between weak and strong coupling. 
This indicates a non-trivial renormalization of the lattice spacing 
of the sort discussed in Section 1. The discrepancy will become 
even more severe as we go to the regime with $g^{2}>>1$ where the 
continuum limit is supposed to lie.  

\section{String Theory Dual of the Confining Branch}
\paragraph{}
The field theory S-duality reviewed in Section 2 predicts that the 
$\beta$-deformed theory with gauge coupling $\tilde{g}^{2}$, 
zero vacuum angle  and 
deformation parameter $\tilde{\beta}=8\pi^{2}i/{\tilde g}^{2}n$ has a 
confining branch.  We will focus on the vacua with 
non-zero chiral VEVs labelled, as in (\ref{svevs}), 
by complex parameters $\tilde{\alpha}_{1}$ and $\tilde{\alpha}_{2}$ 
(with $\tilde{\alpha}_{3}=0$). 
\paragraph{}
In the string theory realization of the $\beta$-deformed theory, 
it is easy to see that field theory S-duality coincides exactly with 
the S-duality of the IIB string. 
Apart from the usual $SL(2,Z)$ action on the complexified string coupling, 
$SL(2,Z)$ transformations act as linear transformations 
on the the RR and NS three-form field strengths $F_{(3)}$ and $H_{(3)}$.    
Equivalently, the complexified three-form field strength $G_{(3)}$ 
transforms with holomorphic weight $(-1,0)$ under $SL(2,Z)$. 
This matches the transformation of the deformation 
parameter $\beta$ under field theory S-duality.  
\paragraph{}
We can now use IIB S-duality to find the string theory dual of the 
confining phase vacuum. Starting with the 
background value for the 
Ramond-Ramond field strength $F_{(3)}$ corresponding to 
$\beta=2\pi/n$, we end up with a non-zero 
background for the Neveu-Schwarz field strength $H_{(3)}$ 
corresponding to the dual theory with 
$\tilde{\beta}=8\pi^{2}i/{\tilde g}^{2}n$.  
Similarly, starting with $m$ D5 branes wrapped on $T^{2}(r_{1},r_{2})$ we end 
up with $m$ NS 5 branes wrapped on the same torus. 
In terms of the S-dual variables we have,
$r_{1}=|\tilde{\alpha}_{1}|(2\pi \tilde{\alpha}')$ 
and $r_{2}=|\tilde{\alpha}_{2}|(2\pi \tilde{\alpha}')$. Finally the $N$ units 
of D3 brane charge are invariant under S-duality and correspond to a constant 
background value of the Ramond-Ramond two-form potential $B_{\rm RR}$ in the 
torus component of the NS5-brane world volume. This configuration 
corresponds to the confining phase vacuum of the dual theory. 
\paragraph{}
It is important to note that, unlike the Higgs branch configuration, 
there is no regime in which which the system is simply described by 
toroidally wrapped branes in flat space. At small 't Hooft coupling 
$\tilde{g}^{2}n<<1$ the deformation corresponding to 
$\tilde{\beta}=8\pi^{2}i/{\tilde g}^{2}n$ becomes large and we cannot ignore 
its effect on the geometry. On the other hand, for $\tilde{g}^{2}n>>1$, 
we have a small deformation of the strongly coupled ${\cal N}=4$ theory. In 
this case, as for the Higgs vacuum, we should find a supergravity dual which 
asymptotes to $AdS_{5}\times\tilde{S}^{5}$  for large values of the 
radial coordinate. In fact we can simply apply the S-duality transformation 
to the geometry describing the 
Higgs phase vacuum derived in the previous Section. 
The S-dual configuration will consist of $m$ NS5 branes wrapped on 
a torus $T^{2}\subset S^{5}$ and located at fixed radial position 
$\bar{r}=\sqrt{r_{1}^{2}+r_{2}^{2}}$ in the near horizon geometry of the 
D3 branes.        
\paragraph{}
The supergravity solution describing the confining phase vacuum 
is obtained by applying the transformation (\ref{2bsdualb}) to the 
Higgs branch solution. Under this transformation, the UV component of 
the solution (\ref{coulomb}), which describes a point 
on the Coulomb branch of the ${\cal N}=4$ theory is self-dual.  
On the other hand the 
geometry near the wrapped branes (\ref{coulomb2}) transforms 
non-trivially and becomes,    
\begin{eqnarray}
d\tilde{s}^{2} \,\,  & = & {\cal Z}_{\rm IR}^{-\frac{1}{2}}
\left(\frac{v_{cr}^{2}+v^{2}}{v^{2}}\right)^{\frac{1}{2}}
\eta_{\mu\nu}
dx^{\mu}dx^{\nu} + {\cal Z}_{\rm IR}^{+\frac{1}{2}}
\left(\frac{v^{2}}{v_{cr}^{2}+v^{2}}\right)^{\frac{1}{2}}
[dx_{4}^{2}+dx_{5}^{2}] \nonumber \\
& & \qquad{} \qquad{} \qquad \qquad{} \,\, + \,\,{\cal Z}_{\rm IR}^{+\frac{1}{2}}
\left(\frac{v_{cr}^{2}+v^{2}}{v^{2}}\right)^{\frac{1}{2}}
[dv^{2}+v^{2}d\Omega_{3}^{2}] \nonumber \\
B_{\rm RR} \,\, & = &\frac{b}{\tilde{\alpha}'}
\left(\frac{v^{2}}{v_{cr}^{2}+v^{2}}\right)  \,\,dx_{4}
\wedge dx_{5} \nonumber \\ 
e^{2\tilde{\phi}} & = & \tilde{g}^{2}_{s}
\left(\frac{v_{cr}^{2}+v^{2}}{v^{2}}\right)  
\nonumber \\ \chi_{4} &= & 
\frac{1}{\tilde{g}_{s}{\cal Z}_{\rm IR}}\,\, 
dx_{0}\wedge dx_{1}\wedge dx_{2}\wedge 
dx_{3} \nonumber \\ 
\label{coulomb3}
\end{eqnarray}
As above we have 
${\cal Z}_{\rm IR} = {\cal R}^{4}/4\pi r_{1}r_{2}v^{2}$ 
where, in terms of the S-dual variables,  
${\cal R}^{4}=4\pi\tilde{g}_{s}N\tilde{\alpha}'^{2}$. As in the 
Higgs branch solution we have the identifications, 
\begin{eqnarray}
x_{4}\sim x_{4}+2\pi r_{1}   & \qquad{} \qquad{} & 
x_{5}\sim x_{5}+2\pi r_{2} 
\label{ident45s}
\end{eqnarray} 
\paragraph{} 
The resulting geometry asymptotes to 
$AdS_{5}\times S^{5}$ near the boundary and undergoes a smooth 
transition to the near-horizon geometry of $m$ flat NS5 branes with background 
$B_{\rm RR}$ given in (\ref{ns5soln2}). 
In particular, for distance from the brane less than the 
cross-over distance,   
\begin{equation}
v_{cr}=\sqrt{\frac{m r_{1}r_{2}}{\tilde{g}_{s}n}}
\label{sparam}
\end{equation}
we enter the infinite throat region of the NS5 brane solution described by 
the metric (\ref{ns5soln3}). Eqn (\ref{ident45s}) implies that 
the coordinates $Y_{4}$ and $Y_{5}$ along the NS5 branes appearing in 
(\ref{ns5soln3}) are identified as, 
\begin{eqnarray}
Y_{4}\sim Y_{4}+2\pi L_{1}   & \qquad{} \qquad{} & 
Y_{5}\sim Y_{5}+2\pi L_{2} 
\label{ident45s2}
\end{eqnarray}   
where, 
\begin{equation}
L_{i}={\cal Z}_{IR}^{\frac{1}{4}}
\left(\frac{v}{v_{cr}}\right)^{\frac{1}{4}}r_{i}=\left(\frac{\tilde{g}_{s}n
\tilde{\alpha}'}{r_{1}r_{2}}\right)^{\frac{1}{2}}r_{i}
\label{convs}
\end{equation}
for $i=1,2$. As before, the patching condition for matching the asymptotic 
forms of the two regions is 
$v_{cr}<<r_{1}$, $r_{2}$. This is satisfied provided we work at 
sufficiently large 't Hooft coupling, $\tilde{g}_{s}n>>m$.  
\paragraph{}
The geometry near the branes coincides with holographic dual of the 
fivebrane worldvolume theory ${\cal T}[M_{s},g_{s}]$. Once again 
we can identify 
the parameters of the five-brane theory in terms of those of the 
four-dimensional gauge theory. We find $g_{s}=4\pi/\tilde{g}^{2}$ and, 
\begin{equation}
M_{s}=\sqrt{\frac{|\tilde{\alpha}_{1}||\tilde{\alpha}_{2}|}{\tilde{g}^{2}n}}
\label{msconf}
\end{equation}
at low energies the fivebrane theory ${\cal T}[M_{s},g_{s}]$ reduces to 
a six-dimensional gauge theory with $G_{6}=1/M_{s}$. This result for $G_{6}$ 
agrees with the field theory result obtained by rewriting the first 
equality in (\ref{table3}) in terms of the S-dual confining phase 
variables.      
\paragraph{}
As for the Higgs branch solution we can attempt to understand our solution 
in terms of an RG flow between four-dimensional superconformal 
field theory in the UV and six dimensional theory in the IR. 
In the UV region $r>>\bar{r}$ where conformal invariance is 
restored, the UV/IR connection provides a relation between the 
radial dependence of the supergravity solution and RG flow in the dual 
field theory. As the deviations from the $AdS_{5}$ geometry become 
significant at $r\sim \bar{r}$ we identify the corresponding energy scale, 
\begin{equation}
\Lambda=
\frac{\bar{r}}{\sqrt{\tilde{g}_{s}N}\tilde{\alpha'}} \, \sim \, 
\frac{1}{\sqrt{\tilde{g}^{2}N}}\,
{\rm max}\left[|\tilde{\alpha}_{1}|,|\tilde{\alpha}_{2}|\right]
\label{escale}         
\end{equation}
This is to be contrasted with the field theory result 
(rewritten in the S-dual variables) 
\begin{equation}
\Lambda_{\rm WC}= \frac{1}{\varepsilon}=
\left(\frac{4\pi}{\tilde{g}^{2}}\right)
{\rm max}\left[|\tilde{\alpha}_{1}|,|\tilde{\alpha}_{2}|\right]     
\label{lamwc34}
\end{equation} 
Once again there is a striking discrepancy between the weak coupling and 
strong coupling results indicating non-trivial renormalisation 
of the lattice spacing in the language of deconstruction. In our 
chosen region of parameters it seems that the effective UV cut-off on the 
six-dimensional theory is not larger than the dynamical scale of the theory 
$M_{s}$. This is quite striking because we are in exactly the regime of 
parameters which should be close to the continuum limit predicted on the 
basis of weak coupling arguments in Section 3. It seems that 
no such continuum limit exists.
\paragraph{}
Despite the above, we still find an interesting result when we take 
the 't Hooft limit $n\rightarrow \infty$, $\tilde{g}^{2}\rightarrow 0$ 
with $\tilde{g}^{2}n$ fixed. The cross-over scale 
$v_{cr}$ remains fixed in this limit as does the dynamical scale $M_{s}$. 
The effective string coupling 
$\exp(\tilde{\phi})$ vanishes uniformly for all $v\geq v_{cr}$. 
The region of the geometry where the coupling remains non-zero is in the 
tube $v<<v_{cr}$. In fact, to find a non-zero effective coupling we need 
to look at the region where $v\sim \tilde{g}_{s}$ as 
$\tilde{g}_{s}\rightarrow 0$. This is precisely the limit discussed in 
Section 4, where the full solution (\ref{ns5soln2}) reduces to 
the near horizon geometry of $m$ NS five-branes with no additional 
fields turned on. The theory in this region is identical to 
Little String Theory and it is completely decoupled from all of the 
other states in the theory. The LST in question is 
compactified on a torus of radii $L_{1}$ and $L_{2}$ given in 
(\ref{convs}) above. As explained at the end of Section 2 we must convert 
these lengths into field theory units to find compactification radii, 
\begin{eqnarray}
R_{1}= \left(\frac{\hat{M}_{s}}{M_{s}}
\right)L_{2}=\frac{\tilde{g}^{2}n}{8\pi^{2} 
|\tilde{\alpha}_{1}|}  & \qquad{} \qquad{} & R_{1}= 
 \left(\frac{\hat{M}_{s}}{M_{s}}
\right)L_{1}=\frac{\tilde{g}^{2}n}{8\pi^{2} 
|\tilde{\alpha}_{2}|}    \nonumber \\
\label{rrr}
\end{eqnarray}
where $\hat{M}_{s}=1/\sqrt{16\pi^{3}\tilde{\alpha}'}$. This agrees exactly 
with the weak coupling formula (\ref{table2}) written in terms of the 
confining phase variables. 
   
\section{The Continuum Limit Revisited}
\paragraph{}
The results of the previous Section establish a modified version of the 
deconstruction conjecture. Summarizing the above, we will  
now state the main result starting from the strongly-coupled confining 
phase of the $\beta$-deformed theory. For brevity, we will now drop all      
tildes on confining phase quantities.     
\paragraph{}
We consider the $\beta$-deformed theory with gauge group $U(mn)$, 
gauge coupling $g^{2}$, deformation parameter $\beta=8\pi^{2}i/g^{2}n$ 
and zero vacuum angle. This theory has 
a moduli space of vacua where the $U(mn)$ gauge symmetry is confined down to a 
$U(m)$ subgroup. We will consider the vacuum state specified by the 
moduli,
\begin{equation}
\begin{array}{ccc} 
\left\langle \frac{1}{N}{\rm Tr}\left[\Phi_{1}^{n}\right]\right\rangle 
= \alpha^{n}_{1} \qquad{} & 
\left\langle \frac{1}{N}{\rm Tr}\left[\Phi_{2}^{n}\right]
\right\rangle=\alpha^{n}_{2} 
\qquad{} &  \left\langle \frac{1}{N}{\rm Tr}\left[\Phi_{3}^{n}\right]
\right\rangle=0 \end{array}
\label{vevsX}
\end{equation}  
\paragraph{}
{\bf Result:} 
In the limit $n\rightarrow\infty$, $g^{2}\rightarrow 0$ with $g^{2}n$, 
$m$, $\alpha_{1}$ and $\alpha_{2}$ fixed, the interacting sector of the 
theory is equivalent to 
Type IIB Little String Theory with low energy gauge group 
$SU(m)$ and string mass-scale, 
\begin{equation} 
M_{s}=\sqrt{\frac{|\alpha_{1}||\alpha_{2}|}{g^{2}n}}
\label{msX}
\end{equation}
compactified to four dimensions on a torus of radii,  
\begin{eqnarray}
R_{1}=\frac{g^{2}n}{8\pi^{2} |\alpha_{1}|} & \qquad{} \qquad{} \qquad{} 
& R_{2}=\frac{g^{2}n}{8\pi^{2}|\alpha_{2}|} \nonumber \\
\label{table2X}
\end{eqnarray}   
with supersymmetry-preserving boundary conditions. 
In addition to the LST degrees 
of freedom, the $\beta$-deformed 
theory contains additional massive and massless states which
are completely decoupled in this limit.
\paragraph{}
In the previous Section we were able to demonstrate this result 
explicitly for the case $g^{2}n>>m$. 
This corresponds 
to the case where the compactification torus is much larger than the Little 
String length scale: $R_{1}R_{2}>>M_{s}^{-2}$. The dual brane configuration 
continues to exist when this restriction is relaxed and the fact that 
fivebrane worldvolume decouples from gravity as the 
asymptotic string coupling goes to zero should be true generally. 
Thus we conjecture that the result holds for all values of $g^{2}n$. 
In the more general case, the approximate supergravity solution obtained above is no longer valid. Provided the weaker condition $g^{2}n>>1/m$ is satisfied, 
the UV behaviour of the theory should still be described by 
classical supergravity 
on $AdS_{5}\times X_{5}$ for some five-manifold $X_{5}$ which is no longer 
a small deformation of the round $S^{5}$. Without further information, 
it would be hard to verify the more general conjecture directly.     
\paragraph{}
We will now review the basic properties of Little String Theory on 
$R^{3,1}\times T^{2}$ and 
the extent to which they are correctly reproduced by our proposed 
deconstruction. 
\paragraph{}     
{\bf Symmetries:} Type IIB Little String Theory has ${\cal N}=(1,1)$ 
super-Poincare symmetry in six dimensions which is broken to 
${\cal N}=4$ super-Poincare symmetry in four dimensions by 
compactification on $T^{2}$. The $U(1)\times U(1)$ symmetry of translations 
on $T^{2}$ is also left unbroken and the corresponding momenta appear as 
central charges in the four-dimensional ${\cal N}=4$ SUSY algebra. 
The low-energy gauge group is $SU(m)$ and the theory also has an 
$SO(4)$ global R-symmetry. 
\paragraph{}
At first sight the symmetries of the $\beta$-deformed theory are 
quite different. At finite $n$ the theory has only ${\cal N}=1$ 
super-Poincare invariance in four-dimensions. The low-energy gauge group 
on the confining branch discussed above is $U(m)$ and the theory has a 
$U(1)^{3}$ R-symmetry broken to 
${\bf Z}_{n}\times {\bf Z}_{n}\times U(1)$ in the 
vacuum (\ref{vevsX}). 
As in other deconstruction scenarios, the idea is that the finite $n$ theory 
corresponds to a lattice regularisation of the higher-dimensional theory 
with the $U(1)\times U(1)$ of translations on $T^{2}$ broken to 
${\bf Z}_{n}\times {\bf Z}_{n}$ corresponding to translations on an 
$n\times n$ lattice. The lattice also breaks the six-dimensional 
${\cal N}=(1,1)$ supersymmetry down to ${\cal N}=1$ in four dimensions. 
This is realised very explicitly in the weakly coupled Higgs branch regime 
of Section 3.1 which corresponds to a very large lattice spacing.  
\paragraph{}
The analysis of the strongly-coupled continuum limit given in the previous 
section showed that, rather than the lattice spacing going to zero, what 
really happens is that an interacting sector of the theory, which 
recovers the full symmetries of LST, decouples completely from the 
rest of the states in the theory. 
It is important to emphasise that the full theory, including the 
extra decoupled states, will respect some but not all of these symmetries. 
As $n\rightarrow \infty$ in the continuum limit, the 
${\bf Z}_{n}\times {\bf Z}_{n}$ R-symmetry of the 
confining phase theory goes over to 
the $U(1)\times U(1)$ symmetry of translations on the torus. On the 
other hand, the full SUSY algebra of LST is only recovered in the 
interacting sector. The 
remaining unbroken $U(1)\simeq SO(2)$ R-symmetry of the $\beta$-deformed  
theory corresponds to a subgroup of the $SO(4)$ R-symmetry of LST.       
The supergravity analysis of the preceeding section shows that the full 
$SO(4)$ symmetry is only recovered in the interacting sector of the theory.    
\paragraph{}
{\bf Spectrum:} The effective action for the 
massless modes of LST compactified on $T^{2}$ is ${\cal N}=4$ SUSY 
Yang-Mills in four dimensions with gauge group $SU(m)$ and gauge coupling 
$G_{4}^{2}=4\pi^{2}M^{2}_{s}R_{1}R_{2}$. Using the conjectured 
identifications (\ref{msX}), we find $G_{4}^{2}=16\pi^{2}/g^{2}n$
The exact effective action for the massless modes of the 
$\beta$-deformed theory on its confining branch is ${\cal N}=4$ SUSY 
Yang-Mills with gauge group $U(m)$ and gauge coupling $g^{2}n$. 
This is consistent with the conjecture because the 
${\cal N}=4$ theory has an exact $SL(2,{\bf Z})$ 
duality which acts as $G_{4}^{2}\rightarrow 16\pi^{2}/G_{4}^{2}$.  
The ${\cal N}=4$ vector multiplet corresponding to the 
central $U(1)\subset U(m)$ should decouple in the proposed continuum limit
\footnote{As the theory contains only adjoint fields, 
the gauge boson of this multiplet is completely decoupled to start with. 
Note however that this is not true of the remaining fields in this 
multiplet at finite $n$.}
\paragraph{}
The agreement of the massless fields means that, locally, 
the vacuum moduli space of the $\beta$-deformed theory is the same as that 
of compactified LST. Globally, however, the two moduli spaces differ at 
finite $n$. The confining branch of the $\beta$-deformed theory 
is the complex orbifold 
${\bf C}^{3}/{\bf Z}_{n}\times{\bf Z}_{n}$. This is a submanifold of a 
larger branch on which the unconfined gauge symmetry is $U(1)^{m}$. 
The full branch is the $m$-fold symmetric product of the orbifold.  
On the other hand the moduli space of LST on 
$T^{2}$ is an $m$-fold symmetric product of $T^{2}\times R^{4}$. 
In fact the two moduli spaces agree in the large-$n$ continuum limit 
by exactly the same mechanism as in the standard deconstruction based 
on a quiver gauge group \cite{AHCK}. In particular one may think of 
${\bf C}^{3}/{\bf Z}_{n}\times{\bf Z}_{n}$ as a cone over an 
$S^{5}/{\bf Z}_{n}\times{\bf Z}_{n}$ base 
which becomes very narrow in the large-$n$ limit. The neighbourhood 
of a point on $C^{3}/{\bf Z}_{n}\times {\bf Z}_{n}$ goes over to 
$T^{2}\times R^{4}$ in 
the large-$n$ limit and the two moduli spaces agree.             
\paragraph{}
LST compactified on the torus also has a spectrum 
of massive BPS states which preserve one half of the supersymmetry algebra. 
These include two towers of Kaluza-Klein states coming from 
compactification of the massless fields in six dimensions.  
In field theory, these states appear in the 
weakly-coupled Higgs branch regime of classical deconstruction. 
Our AdS analysis shows that 
they persist in the strongly coupled theory (at least for $g^{2}n>>1$).   
LST also contains BPS strings in 
six-dimensions which can wind around either cycle of $T^{2}$ thereby 
producing two towers of BPS states in four dimensions. BPS saturated 
strings of the same tension appear as chromomagnetic flux tubes in the S-dual 
dual Higgs description of the confining phase theory. The 
Higgs phase theory reduces to a non-commutative 
six-dimensional gauge theory at low energies and the BPS strings can be 
understood as Yang-Mills instantons embedded in six-dimensions. The 
expected BPS winding modes should correspond to bound states in 
quantum mechanics on the moduli space of instantons. 
\paragraph{}
More generally, the existence of BPS winding modes leads to an 
$SL(2,{\bf Z})\times SL(2,{\bf Z})$ T-duality group for LST compactified to 
four dimensions on a torus. One transformation which is particularly 
interesting coresponds to a simultaneous T-duality on both 
cycles of the torus. 
In the $\beta$-deformed theory this corresponds to 
a transformation which inverts the 't Hooft coupling $g^{2}n$ as, 
\begin{equation}                 
g^{2}n  \rightarrow \frac{16\pi^{2}}{g^{2}n}
\label{stilde}
\end{equation}
As the symmetry relates small and large values of the 't Hooft coupling, 
it is a test of the strong form of the deconstruction conjecture, which 
applies for all values of $g^{2}n$.  
In fact this duality is evident in the low-energy effective action and 
corresponds to an electric-magnetic 
duality transformation of the $U(m)$ ${\cal N}=4$ effective theory of the 
massless modes. If the conjecture is correct this should also be an exact 
duality of the the full theory in the 't Hooft limit.  

\section{An Application}
\paragraph{}
So far we have focussed on deconstructing LST at the origin of its moduli 
space where the low energy gauge symmetry is $SU(m)$. The holographic dual of 
this theory includes a linear background for the dilaton which means that 
the effective string coupling becomes large for small values of the 
radial coordinate. The presence of a strong coupling region makes it hard 
to perform reliable calculations of the observables of LST using the dual 
description. However, LST also has a moduli space of vacua on which 
the low-energy gauge group is broken to $U(1)^{m-1}$. Following \cite{GK}, 
it is possible to define a double-scaling limit of the theory on this 
branch which yields a weakly coupled holographic dual. In this Section, we 
will review the resulting Double-Scaled Little String Theory (DSLST) and 
show how to extend our proposed deconstruction to this case.    
\paragraph{}
As for the theory at the origin, DSLST is defined as a decoupling 
limit of the world volume theory on $m$ D5 branes in 
Type IIB string theory. Before taking any limit, the 
low-energy theory on these branes is a six-dimensional 
$U(m)$ gauge theory which includes 
four real adjoint scalar fields. Each field is an $m\times m$ matrix and, 
with an appropriate choice of normalisation, the eigenvalues specify 
the positions of the branes in their four transverse directions.      
We will consider a configuration where the 
$m$ NS5 branes distributed symmetrically around a circle of radius 
$r_{0}$ in the transverse $R^{4}$. If we 
combine the four real scalar fields into two complex scalars $A$ and $B$, 
this corresponds to choosing VEVs $\langle A \rangle=0$ and 
$\langle B \rangle\sim r_{0}U_{(m)}$. As above, $U_{(m)}$ denotes the 
$m\times m$ clock matrix. 
In this vacuum, the spectrum of the theory includes 
massive W-bosons corresponding to strings
 stretched between the D5 branes. These states have masses, 
\begin{equation}
M^{(ab)}_{W}= \frac{1}{2\pi\alpha'}\, 2r_{0}
\sin\left(\frac{\pi}{m}|a-b|\right)
\label{mw}
\end{equation}         
where $a,b=1,2,\ldots m$ are integers labelling the branes. Note also that 
the global $SO(4)$ symmetry of the theory is spontaneously broken to 
$SO(2)\times {\bf Z}_{m}$ by the scalar VEVs.
\paragraph{}
Double-scaled Little String Theory is obtained by taking the limit 
$g_{s}\rightarrow \infty$, $r_{0}\rightarrow 0$ with the mass scale 
$\bar{M}=1/g_{s}r_{0}$ held fixed. As in the conventional definition of 
LST we also keep the six-dimensional gauge coupling,  
$\hat{G}_{6}=\sqrt{16\pi^{3}\alpha' g_{s}}$ fixed.
As in the basic case, we can reinterpret this limit 
by performing an S-duality transformation (\ref{sdual0}).  
Thus we start from a configuration of $m$ NS5 branes uniformly 
distributed around a circle of radius $r_{0}$. The W-bosons (\ref{mw})
now correspond to D-strings stretched between the NS5 branes. 
The double-scaling limit is now $\tilde{g}_{s}\rightarrow 0$, 
$r_{0}\rightarrow 0$ with 
$\hat{G}_{6}=\sqrt{16\pi^3\tilde{\alpha}'}$ and 
$\bar{M}=\tilde{g}_{s}/r_{0}$ held fixed.      
\paragraph{}
According to \cite{GK}, this limit yields an exact superstring 
background of the form (see also \cite{sf}), 
\begin{equation}
R^{5,1}\times 
\left(\frac{SL(2)}{U(1)}\times \frac{SU(2)}{U(1)}\right)/{\bf Z}_{m}
\label{background}
\end{equation}
where the level of the two cosets appearing in the brackets is equal to 
$m$, the number of NS5 branes. 
The non-compact coset $SL(2)/U(1)$ corresponds to the two-dimensional 
black hole geometry or cigar. 
In an asymptotic region, far from the tip of the cigar, the background has 
a linear dilaton which matches that of the coincident NS5 solution, 
(\ref{ns5soln3}). Importantly, the string coupling is bounded 
above by its value at the tip of the cigar which is controlled by the ratio 
of the two mass scales defined above,  
\begin{equation}
g_{\rm cigar}\sim \frac{\bar{M}}{M_{s}}
\label{gcig}
\end{equation}
Thus, provided we choose this ratio to be small, the background can be 
studied reliably using string perturbation theory. Even better, the 
background (\ref{background}) corresponds to an exactly solvable 
conformal field theory, so that tree-level string theory is 
fully tractable. 
\paragraph{}
It is straightforward to adapt the proposal of the previous section to 
provide a deconstruction of DSLST. The details of the 
necessary modification are given in Appendix B. As in the previous section 
we will drop all tildes on confining branch variables. 
We start from the $\beta$-deformed theory with $\beta=8\pi i/g^{2}n$ in the 
vacuum where the $U(N)$ gauge group, with $N=mn$ is confined down to $U(m)$ 
and the non-zero chiral operators are given by (\ref{vevsX}). The 
low-energy action for the massless degrees of freedom is precisely 
${\cal N}=4$ SUSY Yang-Mills with gauge group $U(m)$. We now 
move onto the Coulomb branch of the low-energy theory by introducing 
a non-zero VEV for the field $\Phi_{3}$ which is written in gauge 
invariant form as,   
\begin{equation}
\left\langle \frac{1}{N}{\rm Tr}\left[\Phi_{3}^{N}\right]
\right\rangle = \mu^{N}
\label{phi3}
\end{equation} 
with all operators of the form ${\rm Tr}[\Phi_{3}^{k}]$ 
having zero VEV for $k<N$. 
Equation (\ref{phi3}) together with (\ref{vevsX}) determines a 
particular vacuum of the theory where the $U(mn)$ gauge group is confined 
down to $U(1)^{m}$. 
\paragraph{}
In the case $g^{2}n>>m$ which corresponds to a large compactification 
torus in string units, we can study the 't Hooft limit using 
AdS duality. The dual geometry again involves $m$ wrapped NS5 branes 
in the Coulomb branch geometry (\ref{coulomb},\ref{bigint}). 
The only difference is that the $m$ branes are now wrapped on $m$ distinct 
tori with small angular seperations on $S^{5}$. As we are ultimately 
interested in a limit where the typical separation, $d\sim 
\mu\tilde{\alpha}'$, between the fivebranes becomes small we can assume 
$d<<v_{cr}$ where 
$v_{cr}$ is the cross-over distance for the fivebrane solution given in 
(\ref{sparam}). Instead of matching onto a solution of $m$ coincident 
NS5 flat branes, we can now patch the solution onto the solution for $m$ 
NS5 branes distributed uniformly around a circle of radius $d$. 
We can then scale $\mu$ in the 't Hooft limit so that the resulting 
decoupled theory is DSLST. The final result is,       
\paragraph{}
{\bf Result:} In the limit $n\rightarrow\infty$, $g^{2}\rightarrow 0$, 
$\mu\rightarrow 0$ with $g^{2}n$, 
$m$, $\alpha_{1}$, $\alpha_{2}$ and $\mu n$ fixed, 
the interacting sector of the 
theory in this vacuum is equivalent to 
double-scaled Little String Theory with low energy gauge group 
$U(1)^{m-1}$ and mass parameters,   
\begin{eqnarray}
M_{s}=\sqrt{\frac{|\alpha_{1}||\alpha_{2}|}
{g^{2}n}} & \qquad{} \qquad{} & \bar{M}=
\frac{2\pi|\alpha_{1}||\alpha_{2}|}{|\mu|n} 
\nonumber \\
\label{finnal}
\end{eqnarray}    
compactified on a torus of radii $R_{1}$ and $R_{2}$ given in 
(\ref{table2X}). In addition to the LST degrees 
of freedom, the $\beta$-deformed 
theory contains additional massive and massless states which
are completely decoupled in this limit. As before our AdS analysis is 
only valid for large 't Hooft coupling $g^{2}n>>m$ and the 
full conjecture which applies for all values of $g^{2}n$ is much 
harder to verify.      
\paragraph{}
Putting everything together we find that, in the t' Hooft large-$n$ limit 
described above, the interacting sector 
of the $\beta$-deformed theory is fully equivalent to the 
IIB string theory background,  
\begin{equation}
R^{3,1}\times T^{2}\times 
\left(\frac{SL(2)}{U(1)}\times \frac{SU(2)}{U(1)}\right)/{\bf Z}_{m}
\label{background2}
\end{equation}
where the IIB squared string-length $\alpha'$ is identified as 
$M_{s}/16\pi^{3}$ and the radii $R_{1}$ and $R_{2}$ of the compactification 
torus are given as in (\ref{table2X}). As above both cosets have 
level $m$, and the dilaton is linear in the asymptotic 
region far from the tip of the cigar. The effective string coupling 
takes its maximum value (\ref{gcig}) at the tip of the cigar. 
Thus if, we also choose $\bar{M}<<M_{s}$, 
the dual string theory is weakly coupled. 
This means that in this corner of parameter space the interacting 
sector of $\beta$-deformed gauge theory can be solved exactly. 
\paragraph{}
The tree-level spectrum and correlation 
functions of DSLST have been studied in detail in \cite{GK} and 
our first task is to understand these results in the context of the 
dual gauge theory. As a preliminary, it is useful to recall some 
basic facts about large-$N$ gauge theories. We begin by considering 
an $SU(N)$ gauge theory containing only adjoint fields in a 
phase where all coloured states are confined. In this case, the 
spectrum will consist entirely of colour-singlet 
glueballs. In the 't Hooft large-$N$ limit, standard counting arguments 
suggest there should be an infinite tower of stable glueballs whose masses 
remain constant as $N\rightarrow \infty$. 
The effective three-point coupling constant governing the interactions of 
these states scales like $1/N$ and thus they become weakly-interacting  
in the large-$N$ limit. The expected spectrum therefore resembles 
that of a closed string theory with effective string coupling constant 
$g^{\rm eff}_{s}\sim 1/N$. The mass scale of the lightest states in the 
spectrum is set by the inverse string tension.     
\paragraph{}
In the present context we are not interested in a standard confining phase, 
but rather in a more exotic phase where a $U(N)$ gauge group is confined 
down to a $U(1)^{m}$ subgroup. As the spectrum includes the gauge bosons 
of the unbroken gauge group there is no mass gap. 
The presence of an unconfined $U(1)^{m}$ 
gauge symmetry also changes the picture because the spectrum can now 
contain states which are electrically or magnetically charged with 
respect to this gauge symmetry. The exact S-duality of the low-energy theory 
interchanges these states and also inverts the 
low-energy gauge coupling: $G^{2}_{4}\rightarrow 16\pi^{2}/G^{2}_{4}$. 
In our string theory construction, the 
electrically charged states correspond 
to D1 branes stretched between the NS5-branes with masses of order 
$M_{W}=M_{s}^{2}/\bar{M}$. There are also 
magnetically charged states corresponding to D3 branes 
stretched between the NS5 branes and wrapped on $T^{2}$ with masses of order 
$g^{2}nM_{W}= R_{1}R_{2}M_{s}^{3}/\bar{M}$. Unlike the colour singlet states, 
there is no reason why these states should be weakly interacting at 
large-$N$. In particular these states couple to the massless $U(1)^{m}$ 
gauge fields with effective gauge coupling $16\pi^{2}/g^{2}n$ which 
remains fixed in the 't Hooft limit.     
\paragraph{}
The above discussion suggests that the large-$N$ behaviour of a partially 
confined phase is quite different from a more conventional phase where the 
whole gauge group is confined. However, as these differences are due to the 
presence of states charged under the unconfined gauge symmetry they should 
disappear in a limit where we decouple the extra charged states. By 
choosing $\bar{M}<<M_{s}$ we ensure that all charged states become 
very massive and we expect the remaining spectrum of colour singlets to 
resemble that of a conventional large-$N$ confining gauge theory but with 
the strength of the residual interactions controlled by the ratio 
$\bar{M}/M_{s}$. 
\paragraph{}       
We can now compare these expectations with the analysis of the string 
theory background (\ref{background}) for $\bar{M}<<M_{s}$ given in 
\cite{GK,AGK}. The first obvious 
point is that we have a dual description in terms of weakly coupled 
closed string theory in this regime with the effective string coupling 
controlled by the ratio $\bar{M}/M_{s}$ which matches our 
gauge theory expectations. The spectrum of string states includes a 
discrete set of states localised at the tip of the cigar. 
These include the expected massless states corresponding to the $m-1$
${\cal N}=(1,1)$ vector multiplets of the low-energy gauge theory.  
The spectrum includes infinite towers of excited string states 
with a Hagedorn density at high energies. The tree-level S-matrix for these 
states exhibits Regge behaviour. These features match the expected 
behaviour of the dual gauge theory discussed above.  
\paragraph{}
There are also other features of DSLST which are quite 
mysterious. In particular, the spectrum of DSLST 
also contains a continuum of states. These states are $\delta$-function 
normalisable plane waves which scatter off the tip of the cigar.   
This is unexpected from the point of view of the dual gauge theory but 
is certainly an artifact of the large-$N$ limit. If we work 
at fixed large $N=mn$ 
and fixed gauge coupling $m/n<<g^{2}<<1$, the dual geometry is no longer a 
cigar of infinite length. Instead, the NS5 brane throat opens out 
into an asymptotically AdS region at some fixed but large distance from 
the tip. There are no such $\delta$-function normalisable states in 
$AdS_{5}$ and a generic plane wave solution in the throat will give rise 
to a non-normalizable solution in the AdS region. In this case it is 
natural to expect that the continuum of DSLST is replaced by a discrete 
spectrum\footnote{Interestingly the mass gap for the continuum modes 
is $M_{s}/\sqrt{m}$ which 
exactly matches the cut-off energy scale $\Lambda$ given in 
(\ref{escale}) above. The latter is the characteristic energy 
for SUGRA states propagating in the transition region between the NS5 brane 
throat and the UV geometry. 
This is consistent with the fact that the continuum 
modes can propagate along the throat and should also be present in the 
transition region. The author thanks Ofer Aharony for explaining this 
point.}. The dual gauge theory provides a natural UV regulator 
for the cigar. Note that the finite-$N$ geometry described above 
contains non-zero RR fields and is no longer an exactly solvable 
string background. However, at large 't Hooft coupling and for a large 
number of fivebranes $m>>1$, the curvature of the background is everywhere 
small and classical supergravity should be reliable. It would be 
interesting to investigate some of the other 
puzzling features of DSLST in this framework such as the 
limits on the existence of off-shell observables and the 
unusual zero-momentum behaviour discovered in \cite{AGK}.                
\paragraph{}
The author would like to thank Ofer Aharony and Tim Hollowood for useful 
discussions. He would also like to thank 
IHES for its hospitality while this work was being completed. 
\section*{Appendix A: The Probe Calculation}
\paragraph{}
As a first step we 
will start by considering a $q$ probe D5 branes each carrying $p$ units of 
D3 brane charge and wrapped on the torus $T(r_{1},r_{2})$, 
defined in (\ref{tr1}),  in a 
geometry of the general form (\ref{coulomb}). If we choose the warp factor 
${\cal Z}_{\rm UV}={\cal R}^{4}/r^{4}$ then the branes are wrapped on a 
$T^{2}$ submanifold of $S^{5}$ located at fixed radial distance 
$r=\bar{r}$ in $AdS_{5}$. However, we will actually find that the stability 
conditions for the probe do not depend on the choice of warp factor.         
\paragraph{}
Initially we will start with the simplest situation where $p$ and $q$ 
are both small so that we can 
neglect the back reaction of the probe on the geometry. 
It is easy to identify the corresponding configuration in the 
dual field theory,  
\begin{equation}
\begin{array}{cc} \langle \Phi_{1}\rangle = 
\alpha_{1}\left(I_{(q)}\otimes U_{(p)}\right) \oplus O_{(N-pq)} &
\qquad{}  
\langle  \Phi_{2}\rangle = \alpha_{2}\left(I_{(q)}\otimes V_{(p)}
\right) \oplus O_{(N-pq)} 
\end{array}
\label{vacprobe}
\end{equation}
and $\langle \Phi_{3}\rangle = 0$ and $O_{(k)}$ denotes the 
$k\times k$ matrix with zero entries. For general values of the 
deformation parameter, this is not a vacuum of the $\beta$-deformed 
theory and yields a non-zero classical potential energy, 
\begin{equation}
V=\frac{4|\alpha_{1}|^{2}|\alpha_{2}|^{2}}{g^{2}}\, pq 
\left|\sin\left(\frac{\beta}{2}-\frac{\pi}{p}\right)\right|^{2} 
\label{pot1}
\end{equation}
The fact that the potential vanishes for $\beta=2\pi/p$ for all values of 
$\alpha_{1}$ and $\alpha_{2}$ indicates that, 
for this value of the deformation parameter, the $\beta$-deformed theory 
has a Higgs branch on which the gauge group is broken as,
\begin{eqnarray}
G=U(N) &\rightarrow &  U(q)\times U(N-pq) 
\end{eqnarray} 
As we are interested in small values of the deformation parameter we will 
work with $1<<p<<N$ and restrict our attention to the case 
$|\beta|\sim 1/p <<1$. In this case the potential simplifies and becomes, 
\begin{equation}
V=\frac{4|\alpha_{1}|^{2}|\alpha_{2}|^{2}}{g^{2}}\, pq 
\left|\frac{\beta}{2}-\frac{\pi}{p}\right|^{2}= 
\frac{4|\alpha_{1}|^{2}|\alpha_{2}|^{2}}{g^{2}}\left(\frac{q\pi^{2}}{p}
-q\pi{\rm Re}[\beta]+ \frac{pq|\beta|^{2}}{4}\right) 
\label{pot2}
\end{equation}
We will now show how the first two terms in this calculation can be 
reproduced (up to numerical constants) in supergravity.   
\paragraph{}
The relevant terms in the Dirac-Born-Infeld action of the probe brane are, 
\begin{eqnarray}
S_{\rm DBI}&=& -\frac{\mu_{5}}{g_{s}}\int \,
d^{6}\xi \left[-{\rm det}\left(G_{\parallel}\right){\rm det}
\left(g_{s}^{-\frac{1}{2}}e^{\frac{\phi}{2}}G_{\perp} + 
2\pi \alpha'{\cal F}\right)\right]^{\frac{1}{2}} \, \, + \, \, \mu_{5}
\int\, \left(C_{6}+2\pi \alpha'{\cal F}\wedge C_{4}\right) \nonumber \\
\label{dbi}
\end{eqnarray}
where $G_{\parallel}$ and $G_{\perp}$ are the pullback metric in the 
$R^{3,1}$ and $T^{2}$ components of the 
D5 worldvolume respectively and ${\cal F}$ is defined in (\ref{gind2}) above.
We choose coordinates $\xi_{A}=x_{\mu}$ for $A=\mu=0,1,2,3$ along 
$R^{3,1}$  and $\xi_{4}=\chi_{1}$, $\xi_{5}=\chi_{2}$ on $T^{2}$ (where 
$\chi_{1}$ and $\chi_{2}$ are defined below equation (\ref{quantum4}) in 
Subsection 6.1). This gives, 
\begin{eqnarray}
{\rm det}\,G_{\parallel}={\cal Z}^{-2} & \qquad{} \qquad{} & 
{\rm det}\,G_{\perp}={\cal Z} 
\label{probemet}
\end{eqnarray}  
\paragraph{}  
As we have $p$ units of D3 brane charge, the flux of ${\cal F}$ is quantized 
according to, 
\begin{equation}
\int_{T^{2}}\, {\cal F}= 2\pi p
\label{quantum2}
\end{equation}
Writing ${\cal F}={\cal F}_{12} d\chi_{1}\wedge d\chi_{2}$, a uniform 
flux density then implies that ${\cal F}_{12}=p/ 2\pi r_{1}r_{2}$. 
One can check that, 
\begin{equation}
\frac{4\pi^{2}\alpha'^{2}{\rm det}{\cal F}}{{\rm det}G_{\perp}}=
\frac{\alpha'^{2}p^{2}}{{\cal Z}r_{1}^{2}r_{2}^{2}}
\label{eval1}
\end{equation}
This quantity is the ratio of the two terms appearing inside the determinant 
in the first term in the DBI action (\ref{dbi}). It is proportional to the 
ratio $\sigma_{3}^{2}/\sigma_{5}^{2}$ where $\sigma_{3}$ and $\sigma_{5}$ 
are the energy density induced by the D3 brane and D5 brane 
charge respectively. If we choose AdS warp factor 
${\cal Z}_{\rm UV}={\cal R}^{4}/r^{4}$ we find,     
\begin{equation}
\frac{\sigma_{3}^{2}}{\sigma_{5}^{2}}\sim 
\frac{p^{2}}{g_{s}N}\,\frac{r^{4}}
{r_{1}^{2}r_{2}^{2}}
\label{eval}
\end{equation}
Thus for a probe brane located at 
$r\sim \bar{r}$, we find that $\sigma_{3}>>\sigma_{5}$ provided 
$g_{s}N<<p^{2}$. In this case the determinant in the DBI action 
(\ref{dbi}) can be expanded as, 
\begin{equation}  
\sqrt{{\rm det}\left(g_{s}^{-\frac{1}{2}}e^{\frac{\phi}{2}}G_{\perp} + 
2\pi \alpha'{\cal F}\right)} \simeq  2\pi\alpha'\sqrt{{\rm det}{\cal F}}
+\frac{{\rm det}G_{\perp}}{4\pi\alpha'\sqrt{{\rm det}{\cal F}}}
\label{expand}
\end{equation}
This expansion reflects the dominance of D3 brane charge over D5 charge 
in our chosen regime of parameters. 
In an approximation where the D5 brane charge is neglected completely, we 
know that the potential experienced by the probe should vanish reflecting 
the vanshing forces between D3 branes. Correspondingly, the 
first term on the right-hand side of (\ref{expand}) is exactly cancelled 
by the contribution of $C_{(4)}$ to the 
Chern-Simons term in the DBI action (\ref{dbi}). In the absence of a 
non-zero backgound for $C_{(6)}$ the first non-cancelling term comes from the 
second term in (\ref{expand}). The resulting contribution to the 
potential energy of the configuration is,    
\begin{equation}
V_{0}= \frac{\mu_{5}}{g_{s}}\int_{T^{2}}\, d^{2}\xi 
\frac{ \sqrt{ {\rm det}G_{\parallel}} 
{\rm det}G_{\perp}}{4\pi\alpha'\sqrt{{\rm det}{\cal F}}}=
\frac{2\pi^{2}\mu_{5}}{\alpha'}\,\frac{qr_{1}^{2}r_{2}^{2}}{g_{s}p}
\label{v0}
\end{equation}
Note that all dependence on the warp factor ${\cal Z}$ has vanished. 
Bearing in mind the identifications 
$r_{1}=|\alpha_{1}|(2\pi \alpha')$ and 
$r_{2}=|\alpha_{2}|(2\pi \alpha')$ we can evaluate $V_{0}$ in terms of the 
dual field theory variables giving, 
\begin{equation}
V_{0}\sim \frac{q  |\alpha_{1}|^{2}|\alpha_{2}|^{2}}{g^{2}p}
\label{v0a}
\end{equation}
This matches the first term on the right-hand side of the field theory 
potential up to numerical constants. 
\paragraph{}
We now introduce a small non-zero background for the complex threeform 
field strength. 
\begin{equation}
G_{(3)}^{+}=G_{(3)}+i \star_{6}G_{(3)}= (i\rho\beta/3g_{s})
{\cal Z}(r)d\omega_{2}
\label{sdg2}
\end{equation}
with,
\begin{equation} 
\omega_{2}=z_{1}z_{2}d\bar{z}_{1}\wedge d\bar{z}_{2} 
+z_{2}z_{3}d\bar{z}_{2}\wedge d\bar{z}_{3} +
z_{3}z_{1}d\bar{z}_{3}\wedge d\bar{z}_{1}  
\label{2form2}
\end{equation}
This result in a non-zero value of the dual 
six form potential. In the geometry (\ref{coulomb}) the relevant equation of 
motion takes the form, 
\begin{eqnarray}
dB_{(6)}-\tau dC_{(6)} & = &  \frac{1}{g_{s}{\cal Z}} G^{+}_{(3)}\wedge 
dx_{0}\wedge dx_{1}\wedge dx_{2} \wedge dx_{3} \nonumber \\
\end{eqnarray}
which yields 
\begin{equation}
C_{(6)}= -\frac{\rho}{3g_{s}}{\rm Re}[\beta\omega_{2}] \wedge 
dx_{0}\wedge dx_{1}\wedge dx_{2} \wedge dx_{3} 
\label{c6}
\end{equation}
\paragraph{}
The six-form potential (\ref{c6}) contributes 
to the potential energy density 
via the $C_{(6)}$ coupling in the DBI action (\ref{dbi}). 
The resulting contribution is,   
\begin{equation}
V_{1}=-\mu_{5}\frac{4\pi^{2}\rho}{9}\,\frac{qr_{1}^{2}r_{2}^{2}}
{g_{s}}\, {\rm Re}[\beta]
\end{equation}
Once again this does not depend on the corresponding warp factor 
${\cal Z}$. In field theory variables we have, 
\begin{equation}
V_{1}\sim \frac{q  |\alpha_{1}|^{2}|\alpha_{2}|^{2}}{g^{2}}\, {\rm Re}[\beta]
\label{v1a}
\end{equation}
This matches the linear term in the field theory potential (\ref{pot2}) up 
to numerical constants. This is already enough to see that the contribution 
$V_{1}$ of the background flux will cancel $V_{0}$, 
and thereby stabilize the probe brane at an arbitrary 
radial position, for an isolated value of the deformation parameter 
${\rm Re}[\beta]$. A more complete calculation would involve determining the 
term of order $|\beta|^{2}$ and fixing the numerical constants as in 
the ${\cal N}=1^{*}$ case of Polchinski and Strassler.   
\paragraph{}
The next step, following \cite{PS}, is to relax the condition $p<<N$. The 
corresponding backreaction of the D3 brane charge carried by the probe can be 
accounted for by modifying the warp factor ${\cal Z}$ appearing in the 
background geometry (\ref{coulomb}). Provided the expansion (\ref{expand}) 
still holds, the calculation goes through exactly as before. In particular 
the leading terms in the potential experienced by the probe do not depend 
on ${\cal Z}$ and our results are unchanged. Next we can go to the 
the case of the Higgs branch vacuum discussed in the text 
which corresponds to $p=n$ and $q=m$ where $N=mn$. In this case the 
corresponding warp factor is given in (\ref{bigint}). To check that the 
probe calculation continues to make sense, we must check the condition  
(\ref{eval1}) for the validity of the expansion (\ref{expand}). 
Using the UV form of the metric ${\cal Z}_{UV}={\cal R}^{4}/r^{4}$ 
for a brane located at $r=\bar{r}$ the condition is satisfied 
provided $g_{s}m<<n$. This condition is the expected one for 
D3 brane charge to dominate over D5 brane charge. 
\paragraph{}
As long as the condition $g_{s}m<<n$ is satisfied the calculation 
of the potential given above is unchanged. The brane 
configuration is stable for a single value of $\beta\sim 1/n$ 
in accordance with field theory expectations. 
A more complete calculation could check that the relevant value is 
actually $\beta=2\pi/n$.          

\section*{Appendix B: Deconstructing DSLST}
\paragraph{}
In this Appendix we fill in some of the details required to adapt the 
deconstruction proposal of Section 5 to the double-scaled case. 
As for the theory at the origin it is convenient to first 
identify the relevant limit of the Higgs branch theory and then use 
S-duality to reinterpret the limit in terms of the confining 
phase theory. 
\paragraph{}
We begin by considering the Higgs phase of the $U(N)$ 
$\beta$-deformed theory with $N=mn$ and $\beta=2\pi/n$. 
We will now consider a particular vacuum of the theory where 
the gauge symmetry is broken to $U(1)^{m}$. In particular we choose the 
scalar VEVs as,  
\begin{equation}
\begin{array}{ccc} \langle \Phi_{1}\rangle = \alpha_{1}I_{(m)}\otimes U_{(n)} &
\qquad{}  
\langle  \Phi_{2}\rangle = \alpha_{2}I_{(m)}\otimes V_{(n)} & \qquad{}   
\langle \Phi_{3}\rangle = \alpha_{3}U_{(m)}\otimes 
V^{\dagger}_{(n)}U^{\dagger}_{(n)} \end{array}
\label{Bvacm}
\end{equation}
where $\alpha_{1}$, $\alpha_{2}$ and $\alpha_{3}$ are complex numbers. 
We can also describe this vacuum in terms of the 
independent gauge-invariant chiral operators which are non-zero. These are, 
\begin{equation}
\begin{array}{ccc} 
\left\langle \frac{1}{N}{\rm Tr}\left[\Phi_{1}^{rn}\right]\right\rangle 
= \alpha^{rn}_{1} \qquad{} & 
\left\langle \frac{1}{N}{\rm Tr}\left[\Phi_{2}^{sn}\right]
\right\rangle=\alpha^{sn}_{2} 
\qquad{} &  \left\langle \frac{1}{N}{\rm Tr}\left[\Phi_{3}^{N}\right]
\right\rangle=(-1)^{N-1}\alpha^{N}_{3} \end{array}
\label{Bvevs}
\end{equation}
where $r,s=1,2,\ldots,m$ are integers. To absorb an irrelevant phase 
we define $\mu=\exp(\pi i(N-1)/N)\alpha_{3}$. Fixing 
these VEVs uniquely specifies the vacuum state. However, in this vacuum 
there are also an 
additional set of non-vanishing chiral operators, 
\begin{equation} 
\left\langle \frac{1}{N}{\rm Tr}
\left[\Phi_{1}^{km}\Phi_{2}^{km}\Phi_{3}^{km}\right]\right\rangle 
= \exp(i\nu_{k})\alpha^{km}_{1} \alpha^{km}_{2} \mu^{km}_{3}  
\label{extraX}
\end{equation}
for integer $k=1,2,\ldots,n$, where $\exp(i\nu_{k})$ is an 
unimportant overall phase. 
\paragraph{}
In addition to breaking the low energy gauge group down to 
$U(1)^{m}$, the scalar VEVs also break the $U(1)^{3}$ R-symmetry of the 
$\beta$-deformed theory down to 
${\bf Z}_{n}\times {\bf Z}_{n}\times {\bf Z}_{m}$. 
The massless fields correspond to $m$ vector multiplets of ${\cal N}=4$ 
supersymmetry. At finite $n$, 
the theory also contains a complicated spectrum of 
massive excitations. If we choose 
$|\mu|<<|\alpha_{1}|$, $|\alpha_{2}|$ then the lightest W-bosons 
have masses, 
\begin{equation}
M^{(ab)}_{W}= 2|\mu|\sin\left(\frac{\pi}{m}|a-b|\right)
\label{Bmw}
\end{equation}
where $a,b=1,2,\ldots,m$ are integers. Each of these states is 
BPS saturated with respect to the low-energy ${\cal N}=4$ supersymmetry. 
Over each of these light states there 
are two towers of states corresponding to the Kaluza-Klein modes 
of a six-dimensional field compactified to four-dimensions on a torus.   
\paragraph{}
At weak coupling, the string theory dual of this vacuum can be worked out 
by a simple modification of the analysis of the 
$\alpha_{3}=0$ case given in \cite{D1}.  As before we have a configuration 
of $N$ D3 branes polarised into $m$ toroidally wrapped D5 branes. For 
$\alpha_{3}\neq 0$, however each D5 brane is wrapped around a particular  
torus $T^{2}_{a}$ for $a=1,2,\ldots, m$. In terms of the complex coordinates 
(\ref{complex2}) the torus $T^{2}_{a}$ is defined by $\rho_{i}=r_{i}=
|\alpha_{i}|(2\pi \alpha')$ for $i=1,2,3$. For the torus 
$T^{2}_{a}$ corresponding to the $a$'th wrapped D5 brane we also have 
the condition, 
\begin{equation}
\psi_{1}+\psi_{2}+\psi_{3}=\frac{2\pi a}{m}
\label{lineq}
\end{equation}
for $a=1,2,\ldots,m$.  The spectrum (\ref{Bmw}) of light W-bosons 
corresponds to that of the lightest 
fundamental strings stretched between the D5 branes for $r_{3}<<r_{1}$, 
$r_{2}$.       
\paragraph{}
For large values of the gauge coupling we can perform an 
S-duality transformation to the dual confining phase theory with 
coupling $\tilde{g}^{2}$ and deformation parameter 
$\tilde{\beta}=8\pi i/\tilde{g}^{2}n$. 
The corresponding gauge invariant vacuum expectation values are, 
\begin{equation}
\begin{array}{ccc} 
\left\langle \frac{1}{N}
{\rm Tr}\left[\tilde{\Phi}_{1}^{rn}\right]\right\rangle 
= \tilde{\alpha}^{rn}_{1} \qquad{} & 
\left\langle \frac{1}{N}{\rm Tr}\left[\tilde{\Phi}_{2}^{sn}\right]
\right\rangle=\tilde{\alpha}^{sn}_{2} 
\qquad{} &  \left\langle \frac{1}{N}{\rm Tr}\left[
\tilde{\Phi}_{3}^{N}\right]
\right\rangle=\tilde{\mu}^{N} \end{array}
\label{sBvevs}
\end{equation}   
with $\tilde{\alpha}_{i}=(\tilde{g}^{2}/4\pi)\alpha_{i}$ 
for $i=1,2$ and $\tilde{\mu}=(\tilde{g}^{2}/4\pi)\mu$. 
The string dual now consists of $m$ NS5 branes wrapped on 
torii $T^{2}_{a}$ for $a=1,2,\ldots, m$. The spectrum of light 
W-bosons corresponds to D-strings stretched between the NS5 branes with 
masses,  
\begin{equation} 
M^{(ab)}_{W}= 2|\tilde{\mu}|
\left(\frac{4\pi}{\tilde{g}^{2}}\right)
\sin\left(\frac{\pi}{m}|a-b|\right)
\label{sBmw}
\end{equation}
\paragraph{} 
At strong coupling, $\tilde{g}^{2}n>>m$, we can find the string dual of 
the vacuum state described above by the methods used in the text. 
The result is again $m$ NS5 branes wrapped on $S^{5}$ in a geometry of the 
form (\ref{coulomb}). The NS5's are now separated in the angular directions 
of $S^{5}$. Importantly, we are interested in the case where 
the typical separation, $r_{0}$, is small compared 
with the thickness of the shell where the supergravity solution is 
well approximated by the near-horizon geometry of $m$ flat NS5 branes. 
In this regime the deformation is equivalent to 
distributing these flat NS5 branes around a circle of radius $r_{0}$ 
in their four transverse directions  
\paragraph{}
Finally we take the continuum limit, Limit \~I as before. 
Thus we take $n\rightarrow \infty$ and 
$\tilde{g}^{2}\rightarrow 0$ with $\tilde{g}^{2}n$, $m$ 
$\tilde{\alpha}_{1}$ and $\tilde{\alpha}_{2}$ fixed. The only new 
feature is that we simultaneously scale $\tilde{\mu}$ so as to 
hold the W-boson masses (\ref{sBmw}) fixed. Thus we simultaneously 
take the limit $|\tilde{\mu}|\rightarrow 0$ with 
$|\tilde{\mu}|n$ fixed. The result of this scaling is DSLST with parameters, 
\begin{eqnarray}
M_{s}=\sqrt{\frac{|\tilde{\alpha}_{1}||\tilde{\alpha}_{2}|}
{\tilde{g}^{2}n}} & \qquad{} \qquad{} & \bar{M}=
\frac{2\pi|\tilde{\alpha}_{1}||\tilde{\alpha}_{2}|}{|\tilde{\mu}|n} 
\nonumber \\
\label{finalB}
\end{eqnarray}

\end{document}